\documentclass[12pt]{article}

\usepackage{vmargin}
\setmarginsrb{3cm}{2cm}{3cm}{3cm}{1cm}{1cm}{1cm}{1cm}
\usepackage{amsmath,bbm}
\usepackage{amssymb}
\usepackage[dvips]{graphicx}
\usepackage{rotating}
\usepackage{psfrag}
\usepackage{fancyhdr}
\newcommand{\sspe}{< P s,s>}
\newcommand{\sspb}{\overline{< P s^{\prime},s >}}
\newcommand{\ssp}{< P s,s^{\prime} >}
\newcommand{\phys}{\ \ =_{ \left.\right. _{\left.\right._{\!\!\!\!\!\!\!\!\!\!\!\!\!phys}}} \ \ }
\newcommand{\Pe}{<P_{\epsilon}s, s^{\prime}>}
\newcommand{\p}{<s, s^{\prime}>_p}

\newcommand{\Pes}{<P^{\sigma}_{\epsilon} s, s^{\prime}>}
\newcommand{\C}{\mathbb{C}}
\newcommand{\va}{\scriptscriptstyle}
\newcommand{\vani}{\scriptstyle}
\newcommand{\R}{\mathbb{R}}

\newcommand{\Hp}{{\cal H}_{phys}}
\newcommand{\Hk}{{\cal H}_{kin}}

\newcommand{\Ha}{{\cal H}_{aux}}
\newcommand{\PP}{{ P}}
\newcommand{\So}{{\hat {\cal S}}}

\newcommand{\M}{{\cal M}}

\newcommand{\Seg}{\Sigma^{\Gamma\Gamma^{\prime}}_{\epsilon}}

\newcommand{\be}{\nopagebreak[3]\begin{equation}}
\newcommand{\ee}{\end{equation}}
\newcommand{\ba}{\nopagebreak[3]\begin{eqnarray}}
\newcommand{\ea}{\end{eqnarray}}

\author{Karim Noui\thanks{noui@gravity.psu.edu} \hspace{0.2cm}and Alejandro Perez \thanks{perez@gravity.psu.edu}\\ \\
{\it Center for Gravitational Physics and Geometry} \\
{\it Pennsylvania State University} \\
{\it University Park, PA 16802, USA}}
\title{\bf Three dimensional loop quantum gravity: \\
 physical scalar product \\
and spin foam models}
\date{\today}

\begin{document}

\maketitle

\begin{abstract}
{ In this paper, we address the problem of the dynamics in three
dimensional loop quantum gravity with zero cosmological constant. We
construct a rigorous definition of Rovelli's generalized projection
operator from the kinematical Hilbert space---corresponding to the
quantization of the infinite dimensional kinematical configuration space of the theory---to the physical Hilbert space. In
particular, we provide the definition of the physical scalar product
which can be represented in terms of a sum over (finite) spin-foam
amplitudes. Therefore, we establish a clear-cut connection between
the canonical quantization of three dimensional gravity and spin-foam
models. We emphasize two main properties of the result: first that no cut-off in the kinematical degrees of
freedom of the theory is introduced (in contrast to standard `lattice'
methods), and second that no ill-defined sum over spins (`bubble' divergences)
are present in the spin foam representation.}
\end{abstract}

\newpage

\section*{1. Introduction}
The goal of the {\em spin foam} approach \cite{a18} is to construct a
mathematical well defined notion of path integral for loop quantum
gravity as a devise for computing the `{\em dynamics}' of the
theory. By `{\em dynamics}' here we mean the characterization of the
kernel of the quantum constraints of the theory given by the
quantization of the classical constraints of general relativity
expressed in connection variables, namely: \be G_i = D_a E^a_i,\ \ \ \
V_a = E^b_i F^i_{ab} \ \ \ and \ \ \ \ S = \epsilon^{ij}{}_k E_i^a
E_j^b F^k_{ab} \;.  \ee { These constraints are respectively the Gauss
constraints $G_i$, the vectorial constraints $V_a$ and the scalar
constraint $S$ (with zero cosmological constant). Loop quantum gravity
aims at characterizing the space of solutions of quantum Einstein's
equations represented, in the canonical framework, by the previous set
of constraints (for recent reviews see \cite{ash10,c9,threv} and the beautifully
new book by Rovelli \cite{book}; for an extensive and deep description of the
mathematical structure of the theory see the book of Thiemann \cite{bookt}).
Solutions of Gauss and vectorial
constraints are well understood and are described in terms of
spin-networks. They form a Hilbert space ${\cal H}_{kin}$ whose scalar
product is defined by the Ashtekar-Lewandowski (A.L.) measure and is
denoted $<,>$ in the following. The scalar constraint is much more
involved to implement and is still problematic.}

Spin foam models have been studied as an attempt to give an explicit
construction of the generalized projection operator $\PP$ from $\Hk$
into the kernel of the quantum { scalar} constraint. The resolution of
constraint systems using the generalized projection operator $P$ (also called
{\em rigging map}) has been studied in great generality by many authors.
For references and a description of the main ideas and references see \cite{rac0}.
The technique was used to solve the diffeomorphism constraint in loop quantum
gravity in \cite{ash5}. The idea that the generalized projection operator $P$
onto the kernel of the scalar constraint can be represented as a sum over spin
foam amplitudes was introduced in \cite{reis6,c2}. In this latter work the
authors investigate a regularization of the
formal projector into the kernel of the scalar constraint $S(x)$,
given by
\begin{equation}
\label{P} P=``\prod_{x \in \ \Sigma}\delta(\So
(x))"=\int D[N] \ {\rm exp}(i\int \limits_{\Sigma} N \hat {\cal
S})\;,
\end{equation}
was presented.
One can also define the notion of path integral for gravity
as a lattice discretization of the formal path
integral for GR in first order variables
\begin{equation}\label{path}
P=\int \ D[e]\  D[A]\ \mu[A,e]\ {\rm exp}\left[ i S_{\va GR}(e,A) \right]
\end{equation}
\vskip-.1cm \noindent where the formal measure $\mu[A,e]$ must be determined
by the Hamiltonian analysis of the theory. The issue of the measure
\cite{myo} has been so far neglected in the definition of spin foam models of
this type.
On the other hand, no much progress in developing the formal
definition of (\ref{P}) proposed in \cite{reis6,c2}  has been attained in part because of
the ambiguities present in the construction of the scalar constraint, and also because of open questions regarding the finiteness of the
proposed regularization. To see this in more detail let us briefly describe
the main idea introduced by Rovelli and Reisenberger in the context of loop
quantum gravity. In this
case one is concerned with the definition of (\ref{P}) in the
canonical framework.  Given two {\em spin network} states $s,s^{\prime}$ the physical
scalar product $<s,s^{\prime}>_{ph}:=<Ps,s^{\prime}>$ can be formally
defined by \vskip-.3cm
\begin{equation} \label{vani} \left<Ps,
s^{\prime}\right>=\int {D}[N] \sum \limits^{\infty}_{n=0}
\frac{i^{n}}{n!}<\left[\int \limits_{\Sigma} N(x) \hat {\cal
S}(x)\right]^n \ s, s^{\prime}>,
\end{equation}
\vskip-.1cm \noindent where the exponential in (\ref{P}) has been
expanded in powers. From early on, it was realized that smooth
loop states are naturally annihilated (independently of any
regularization ambiguity) by $\hat {\cal S}$ \cite{jac,c8}
\footnote{This set of states is clearly too small to represent the physical
Hilbert space (e.g., they span a zero volume sector).}.  { In fact, one can
show that} $\hat S$ acts only on {\em spin network} nodes. Generically, it
does so by creating new links and nodes modifying in this way the underlying
graph of the {\em spin network} states \cite{th2,qsd}.  The action of $\So$ can be visualized
as an `interaction vertex' in the time evolution of the node (Figure
\ref{pilin}).  Therefore, each term in the sum (\ref{vani}) represents a
series of transitions--given by the local action of $\hat {\cal S}$ at {\em
spin network} nodes--through different {\em spin network} states interpolating
the boundary states $s$ and $s^{\prime}$ respectively. They can in fact be
expressed as sum over `histories' of {\em spin networks} that can be pictured
as a system of branching surfaces described by a 2-complex. In this picture,
links `evolve' to form 2-dimensional faces (that inherit the corresponding
spin label) and nodes evolve into 1-dimensional edges (inheriting
intertwiners). The action of $\hat {\cal S}$ generates vertices were new nodes
are created and a transition to a different {\em spin network} takes
place. Every such history is a {\em spin foam} (see Figure \ref{pilin}).

Before even considering the issue of convergence of this series,
the problem with this definition is evident: every single term in
the sum is a divergent integral! Therefore, this way of presenting
{\em spin foams} has to be considered as formal until a well
defined regularization of (\ref{P}) is provided. Possible regularization
schemes are discussed in \cite{c2} although they have not been implemented
in concrete examples.

The underlying discreteness discovered in loop quantum gravity is
crucial: in {\em spin foam} models the functional integral for
gravity is replaced by a sum over amplitudes of combinatorial
objects given by foam-like configurations ({\em spin foams}). The
precise definition was first introduced by Baez in \cite{baez5}. A
{\em spin foam} represents a possible history of the gravitational
field and can be interpreted as a set of transitions through
different quantum states of space. Boundary data in the path
integral are given by polymer-like excitations ({\em spin
network} states) representing $3$-geometry states in loop quantum
gravity.

Pure gravity in three dimensions is a well studied example of integrable
system that can be rigorously quantized (for a review see \cite{carlip}). The
reason for that is the fact that GR in three dimensions does not have local
degrees of freedom. The degrees of freedom are topological and therefore
finitely many. { Different quantization schemes have been explored and one can
say that we have a well understanding of three dimensional quantum
gravity}. From our perspective three dimensional gravity is taken as an toy
model for the application of quantization techniques that are expected to be
applicable in four dimensions (see \cite{phoe} for a novel proposal and an
account and a discussion technical difficulties arising in four dimensions).
In this sense we want to quantize the theory
according to Dirac prescription which implies having to deal with the
infinitely many degrees of freedom of a field theory at the kinematical level,
i.e., we want to quantize first and then reduce at the quantum level.
This is precisely the avenue that is explored by loop quantum gravity in four
dimensions.

{The next section is devoted to a brief review of the canonical
formulation of three dimensional gravity \`a la loop gravity. However,
our approach is a bit different than the usual loop gravity
description for we treat the vectorial constraints and the scalar
constraint together in the curvature constraint $F_{ab}=0$. The Gauss
constraint still exists and the kinematical Hilbert space ${\cal
H}_{kin}$ is defined by the set of its solutions endowed with the
A.L. measure. In section 3, we address the problem of the dynamics,
i.e. we characterize the physical Hilbert space ${\cal H}_{phys}$ by
providing a regularization of the Rovelli's generalized projection
operator. The physical scalar product can be represented a a sum over
spin foams whose amplitudes coincide---when restricted in a suitable way---
with the definition of the covariant
path integral on a triangulation: the Ponzano-Regge model. Then, we
propose a basis for the physical Hilbert space.  Finally, we end up
with a discussion on the possible generalization of these results (the most natural case being
the case of a non-vanishing cosmological constant) and on the possibility to adapt
this method to the four dimensional framework.}

\section*{2. Canonical three dimensional gravity}

The theory we are interested in is three dimensional
gravity in first order { formalism}. { The space-time $\cal M$ is a three dimensional oriented smooth manifold} and the action is simply given
by
\be
S[e,\omega]=\int_{\cal M} {\rm Tr}[{e} \wedge F({\omega})]
\ee
where $e$ is the triad, i.e. a Lie algebra ($\mathfrak g$) valued $1$-form, $F(\omega)$ is the curvature of the
three dimensional connection $\omega$ and $Tr$ denotes a Killing form on $\mathfrak g$. For simplicity we will concentrate on Riemannian gravity
so the previous fields should be thought as defined on $SU(2)$ principal bundle over $\M$ { and the Lie algebra is $\mathfrak g =su(2)$}.
We assume the space time topology to be $\M = \Sigma \times \R$ where $\Sigma$
is a Riemann surface of arbitrary genus.

\subsection*{2.1. Phase space, constraints and gauge symmetries}

Upon the standard 2+1 decomposition,
the phase space in these variables is parametrized by the pull back to $\Sigma$ of $\omega$ and
$e$. In local coordinates we can express them in terms of the 2-dimensional connection
$A_a^{i}$ and the triad field $E^b_j=\epsilon^{bc} e^k_c \eta_{jk}$ where $a=1,2$ are space coordinate indices and
$i,j=1,2,3$ are $su(2)$ indices. The symplectic structure is defined by
\[\{A_a^{i}(x), E^b_j(y)\}=\delta_a^{\, b} \; \delta^{i}_{\, j} \; \delta^{(2)}(x,y).\]
Local symmetries of the theory are generated by the first class constraints
$D_b E^b_j \simeq 0$ and $F_{ab}^i(A) \simeq 0$. { More precisely, if we smear them out with arbitrary test fields $\alpha$ and $\lambda$, the constraints read:}
\be
\label{gauss} G[\alpha, A,E]=\int \limits_{\Sigma}\alpha^j D_b
E^b_j=0\;\;\; \text{and} \;\;\; C[\lambda, A]=\int
\limits_{\Sigma}\lambda_j F_{ab}(A)^j=0.
\ee
{ In the sequel, we will assume that the test fields $\alpha$ and $\lambda$ do not depend on the phase space variables.} The Gauss constraint $G[\alpha, A,E]$ generates infinitesimal $SU(2)$ gauge
transformations
\be \delta_{\alpha} A^{i}_a=\{A_{i}^a, G[\alpha, A,E]\}=
(D_a \alpha)^{i}, \ \  \delta_{\alpha} E_{i}^a=\{E_{i}^a, G[\alpha,
A,E]\}=\alpha^{k}E^{j a} \epsilon_{ijk},
\ee
while the curvature constraint $C[\lambda, A]$ generates the following transformations:
\be
\delta_{\lambda} A^{i}_a= \{A^{i}_a, C[\lambda, A]\} =0, \ \
\delta_{\lambda} E_{i}^a=\{E_{i}^a, C[\lambda, A]\}= \epsilon^{ac}D_c\lambda
\ee
As it is well known, diffeomorphisms are contained in the
previous transformations. More precisely, { given a vector field $v=v^a \partial_a$ on the surface $\Sigma$}, one defines the parameters $\alpha^i (v)= v^a A_a^i$ and $\lambda_i(v)=\epsilon_{ab}E^a_i v^b$ and we have
\be
({\cal L}_v
A)_a^i \simeq \delta_{\alpha(v)} A_{i}^a, \ \ ({\cal L}_v
E)^a_i \simeq \delta_{\alpha(v)} E^{i}_a + \delta_{\lambda(v)} E^i_a
\ee
where { ${\cal L}_v$ is the Lie derivative operator along the vector field $v$} and $\simeq$ denotes weak equalities, i.e., valid on the constraint surface. { Note that gauge transformations are equivalent to diffeomorphisms if and only if we restrict the $E$-field to be non-degenerate.}

\subsection*{2.2. Kinematical Hilbert space}

In analogy with the four dimensional case we follow Dirac's
procedure and in order to quantize the theory we first find a
representation of the basic variables in an auxiliary Hilbert
space $\Ha$. The basic functionals of the connection are
represented by the set of holonomies along paths $\gamma \subset
\Sigma$. Given a connection $A$ and a path $\gamma$, one defines the
holonomy $h_{\gamma}[A]$ by
\be
\label{hol}h_{\gamma}[A]=P \exp\int_{\gamma} A \;.
\ee
{ As for the triad, its associated basic variable} is
given by the smearing of $E$ along co-dimension 1 surfaces. One
promotes these basic variables to operators acting on an auxiliary
Hilbert space where constraints are represented. The physical
Hilbert space is defined by those `states' that are annihilated by
the constraints. As these `states' are not normalizable with
respect to the auxiliary inner product they are not in $\Ha$ and
have to be regarded rather as distributional.

The auxiliary Hilbert space is defined by the Cauchy completion of
the space of cylindrical functionals ${Cyl}$, on the space of
(generalized) connections $\bar {\cal A}$\footnote{A generalized
connection is a map from the set of paths $\gamma \subset \Sigma$ to
$SU(2)$. It corresponds to an extension of the notion of holonomy
$h_{\gamma}[A]$ introduced above.}. The space $Cyl$ is defined as
follows: any element of $Cyl$, $\Psi_{\Gamma,f}[A]$ is a functional
of $A$ labeled by a finite graph $\Gamma \subset \Sigma$ and a
continuous function $f: SU(2)^{N_\ell({\Gamma})}\rightarrow \C$
where $N_\ell({\Gamma})$ is the number of links of the graph
$\Gamma$. Such a functional is defined as follows
\be \label{cyl}
\Psi_{\Gamma,f}[A]=f(h_{\gamma_1}[A],\cdots,h_{\gamma_{N_{\ell}(\Gamma)}}[A])
\ee
where $h_{\gamma_i}[A]$ is the holonomy along the link $\gamma_i$ of the graph $\Gamma$. { If one considers a new graph $\Gamma'$} such that $\Gamma \subset \Gamma^{\prime}$, then any cylindrical
function $\Psi_{\Gamma,f}[A]$ trivially corresponds to a cylindrical
function $\Psi_{\Gamma^{\prime},f^{\prime}}[A]$ \cite{ash3}.

For example, let us consider the path $\alpha(t)$ given by
$\alpha: [0,3] \rightarrow \Sigma$. We define $\Gamma$ to be the
graph given by the single link $\gamma=\{ \alpha(t)\ for \ t \in
[0,2]\}$ and by the two nodes $\alpha(0)$ and $\alpha(2)$. The graph
$\Gamma^{\prime}$ is defined by the three links
$\gamma_i=\{\alpha(t)\ for\  t\in [i-1,i]\}$ and the four nodes
$\alpha(0),\alpha(1),\alpha(2)$ and $\alpha(3)$. Clearly
$\Gamma \subset \Gamma^{\prime}$. Therefore, given
$\Psi_{\Gamma,f}[A]=f(h_\gamma[A])$, we can construct
$\Psi_{\Gamma^{\prime},f^{\prime}}[A]=f^{\prime}(h_{\gamma_{1}}[A],h_{\gamma_{2}}[A],h_{\gamma_{3}}[A])$ such that $\Psi_{\Gamma,f}[A]= \Psi_{\Gamma^{\prime},f^{\prime}}[A]$ by choosing
$f^{\prime}(h_{\gamma_{1}}[A],h_{\gamma_{2}}[A],h_{\gamma_{3}}[A])=f(h_{\gamma_{1}}[A]h_{\gamma_{2}}[A])$.

In general, for any two cylindrical functions
$\Psi_{\Gamma_1,f}[A]$ and $\Psi_{\Gamma_2,g}[A]$, the inner
product is defined by the Ashtekar-Lewandowski measure \ba
\label{innerk}\nonumber &&
\mu_{AL}(\overline{\Psi_{\Gamma_1,f}[A]}\Psi_{\Gamma_2,g}[A])=<\Psi_{\Gamma_1,f},\Psi_{\Gamma_2,g}
> \\ &&:=\int \prod \limits_{i=1}^{N_{\ell_{\Gamma_{12}}}} dh_i \overline{f(h_{\gamma_1},\cdots,h_{\gamma_{N_\ell(\Gamma_{12})}})} g(h_{\gamma_1},\cdots,h_{\gamma_{N_\ell(\Gamma_{12})}})
\ea where $dh_i$ corresponds to the invariant $SU(2)$-Haar
measure, { $\Gamma_{12} \subset \Sigma$ is a graph containing both
$\Gamma_1$ and $\Gamma_2$, and we have used the same notation $f$
(resp. $g$) to denote the extension of the function $f$ (resp.
$g$) on the graph $\Gamma_{12}$}. The auxiliary Hilbert space
${\cal H}_{aux}$ is defined as the Cauchy completion of $Cyl$
under (\ref{innerk}).

The (generalized) connection is quantized  by promoting the holonomy (\ref{hol}) to an operator acting by multiplication in ${\cal H}_{aux}$ as follows:
\be
\widehat{h_\gamma[A]} \Psi[A] \; = \; h_\gamma[A] \Psi[A]\;.
\label{ggcc}
\ee
It is easy to check that the quantum holonomy is a self adjoint operator in ${\cal H}_{aux}$. The triad is promoted to a self adjoint operator valued distribution that acts as a
derivation,  namely:
\be
\hat{E}^j_a=-i \ell_p \epsilon_{ab} \eta^{kj} \frac{\delta}{\delta A^k_b} \;\;,
\ee
where $\ell_p=\hbar G$ is the Planck length in three dimensions. In terms of
the triad operator we can construct geometric operators corresponding to the
area of regions in $\Sigma$ or the length of curves \cite{th6,c13,c14}.
So far we have not specified the space of graphs that we are considering.

%

The Gauss constraint (\ref{gauss}) can be defined in terms of the
basic variables introduced above. It generates gauge
transformations whose action on $Cyl$ transforms the holonomy as follows
\be \label{gg}
h_{\gamma}[A] \longmapsto  g_s h_{\gamma}[A]g_t^{-1}
\ee
where $g_s,g_t\in SU(2)$ are group elements associated to the {\em
source} and {\em target} nodes of $\gamma$ respectively. The
so-called {\em kinematical} Hilbert space $\Hk \subset \Ha$ is
defined by the states in $\Ha$ which are gauge invariant and hence
in the kernel of the Gauss constraint.

One can introduce an orthonormal basis of states in $\Ha$ using
$SU(2)$ harmonic analysis. Namely, any (Haar measure) square integrable function $f:SU(2)\rightarrow \C$
can be expanded in terms of unitary irreducible representations of $SU(2)$
\be \label{PW}
f(h)=\sum \limits_j \ f_j \ \stackrel{j}{\Pi}(h) \; ,
\ee
where the Fourier component $f_j=\int dh  \; \overline{\stackrel{j}{\Pi}(h)} f(h)$ can be viewed as an element of $j^* \otimes j$ (we use the same notation for the representation and its associated vector space). The straightforward generalization of this { decomposition} to functions $f:SU(2)^{N}\rightarrow \C$
allows us to write any cylindrical function (\ref{cyl}) as
\be
\Psi_{\Gamma,f}[A]=\sum\limits_{j_1\cdots j_{N_\ell}} f_{j_1 \cdots j_{N_\ell}}
\stackrel{j_1}{\Pi}\!(h_{\gamma_1}[A])\cdots \stackrel{j_{N_\ell}}{\Pi}\!(h_{\gamma_{N_{\ell}(\Gamma)}}[A]),
\label{17}
\ee
where Fourier components $f_{j_1 \cdots j_{N_\ell(\Gamma)}}$ are elements of $(\otimes_{i=1}^{N_\ell(\Gamma)} j_i) \otimes (\otimes_{i=1}^{N_\ell(\Gamma)} j_i)^*$.

We can write any element of $Cyl$ as a linear combination of the tensor product of
$N_{\ell}(\Gamma)$ $SU(2)$ irreducible representations. Using the definition of the scalar
product and (\ref{PW}) one can easily check that these elementary cylindrical
functionals are orthogonal.

\begin{figure}[h]
\centering {\includegraphics[width=10cm]{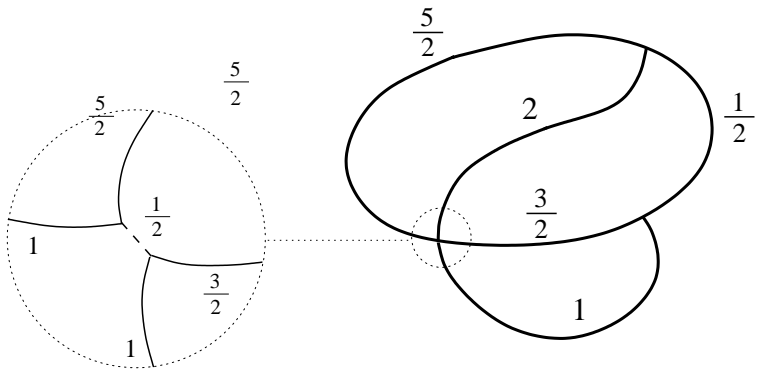}}
\caption{\small Illustration of a spin-network state. At 3-valent nodes the intertwiner is uniquely specified by the corresponding spins. At 4 or higher
valent nodes an intertwiner has to be specified. Choosing an
intertwiner corresponds to decompose the $n$-valent node in terms
of $3$-valent ones adding new virtual links (dashed lines) and
their corresponding spins. This is illustrated explicitly in the
figure for the $4$-valent node.}\label{spinn}
\end{figure}

Gauge transformations (\ref{gg}) generated by the Gauss constraint { induce a gauge action on Fourier modes} which simply reads:
\[\stackrel{j}{\Pi}\!(h)\rightarrow\stackrel{j}{\Pi}\!(g_s)\stackrel{j}{\Pi}\!(h)\stackrel{j}{\Pi}(g^{-1}_t).\]
A basis of gauge invariant functions can then be constructed by contracting
the tensor product of representation matrices in Equation (\ref{17}) with
a $su(2)$-invariant tensors or $su(2)$-intertwiners. If we select an orthonormal basis of intertwiners
$\iota_n \in {\rm Inv}[j_1\otimes j_2\otimes\cdots\otimes j_{\va N_{\ell}}]$
where $n$ labels the elements of the basis we can write a basis of gauge
invariant elements of $Cyl$ called the {\em spin network basis}. Each spin network is
labeled by a graph $\Gamma \subset \Sigma$, a set of spins $j_{\gamma}$
labeling links $\gamma$
 of the graph $\Gamma$  and a set of intertwiners $\iota_n$ labeling nodes $n$ of the graph $\Gamma$, namely:
\begin{equation}
s_{\va \Gamma, \{j_{\ell}\},\{\iota_{n}\}}[A]=\bigotimes_{n \in
\Gamma} \ \iota_n\ \bigotimes_{\gamma \in \Gamma}\  \stackrel{j_{\gamma}}{\Pi}(h_{\va
\gamma}[A]) \;.
\end{equation}
We have represented a spin network state in Figure (\ref{spinn}). In order to lighten notations, we will omit indices (the graph, representations and intertwiners) labeling spin-networks in the sequel.

\section*{3. Dynamics and spin foams}

In this section we completely solve $2+1$ Riemannian { quantum} gravity by
providing a regularization of the Rovelli's generalized projection operator
$P$. { This allows us to construct the physical scalar product: we present its
spin foam representation and its relationship with the Ponzano-Regge model.}
Finally, we provide a complete characterization of the physical Hilbert space
$\Hp$ which is analogous to the one provided in \cite{tti}. Moreover, the
spin foam representation allow us to find a basis in $\Hp$.


\subsection*{3.1. Regularization of P}

In this section we introduce the regularization of the generalized projection operator
$P$ and show how the matrix elements $<Ps,s^{\prime}>$ admit a spin foam state
sum representation that is independent of the regularization and it is not
based on any auxiliary structure such as a triangulation or cellular
decomposition when the regulator is removed.

We start with the formal expression
\begin{equation}
P =``\prod_{x \in \ \Sigma} \delta(\hat F(A))"=\int D[N] \ {\rm
exp}(i\int \limits_{\Sigma} {\rm Tr}[ N \hat {F}(A)])
\;,\label{ppp}
\end{equation}
where $N\in su(2)$.  We now introduce a regularization of
(\ref{ppp}). We will give a definition of $P$ by providing a
regularization of its matrix elements $<Ps,s^{\prime}>$ for any
pair of spin network states $s,s^{\prime} \in \Hk$. Let's denote
by $\Gamma$ and $\Gamma^{\prime}$ the graphs on which $s$ and
$s^{\prime}$ are defined respectively. We introduce an arbitrary
cellular decomposition of $\Sigma$ denoted $\Sigma^{\Gamma
\Gamma^{\prime}}_{\epsilon}$, where $\epsilon\in \R$, such that:
\begin{enumerate}
\item The graphs $\Gamma$ and $\Gamma^{\prime}$ are both contained in the
  graph defined by the union of $0$-cells and $1$-cells in $\Seg$.
\item For each individual 2-cells (plaquette) $p$ there exist a
ball ${\cal B}_{\epsilon}$ of radius $\epsilon$---defined using
the local topology---such that $p\subset{\cal B}_{\epsilon}$.
\end{enumerate}
Consequently all 2-cells shrink to zero when $\epsilon \rightarrow
0$.


Based on the cellular decomposition $\Seg$ we can now define
$\ssp$ by introducing a regularization of the right hand side of
(\ref{ppp}). Given two spin networks $s$ and $s^{\prime}$ based on
the graphs $\Gamma$ and $\Gamma^{\prime}$, we order the set of
plaquettes $p^i\in \Sigma^{\Gamma\Gamma^{\prime}}_{\epsilon}$ for
$i=1,\cdots N^{\epsilon}_p$ where $N^{\epsilon}_p$ is the total
number of plaquettes for a given $\epsilon$. We define the
physical scalar product between $s$ and $s^{\prime}$ as
\be\label{final1}
 \p=\ssp := \lim_{\epsilon\rightarrow 0} \ \  \sum_{j_{\va p^i}}
 (2j_{\vani p^i}+1)\ <\prod_{p^i}
 \chi_{j_{\va p^i}}({U}_{p^i})\ s, s^{\prime}>, \ee
where the sum is over all half-integers $j_{\va p^i}$ labelling
each plaquette, ${U}_{p^i}$ is the holonomy around $p^i$ (based on
an arbitrary starting point) and $\chi_{j_{\va p^i}}({U}_{p^i})$
is the trace in the $j_{\va p^i}$ representation.

Notice that each term in the previous sum is the matrix element of
a self adjoint operator in $\Hk$ and therefore well defined. The
question is whether the previous expression is well defined. As we
will see in the sequel (see remarks at the end of this section),
for a fixed value of $\epsilon$, the sum inside the limit is
convergent for any Riemann surface of genus $g\ge 2$. We also show
in Section 3.2.1. that the result is independent of the
regulator $\epsilon$ and therefore the limit exists trivially. In
the case of the sphere $g=0$ and the torus $g=1$ the previous sum
is divergent; we discuss the regularization of the physical inner
product in these special cases in Section \ref{}.

Let us now give the motivation for the definition of the inner
product above. We consider a local patch $U\subset \Sigma$ where
we choose the cellular decomposition to be square with cells of
coordinate length $\epsilon$. In that patch, the integral in the
exponential in (\ref{ppp}) can be written as a Riemann sum
\[F[N]=\int\limits_{U} {\rm Tr}[ N {F}(A)]=
\lim_{\epsilon\rightarrow 0}\ \sum_{p^i} \epsilon^2 {\rm
Tr}[N_{p^i} F_{p^i}],\] where $p^i$ labels the $i^{th}$ plaquette
and $N_{p^i}\in su(2)$ and $F_{p^i}\in su(2) $ are values of
$N^j\tau_j$ and $\tau_j \epsilon^{ab}F^{j}_{ab}[A]$ at some
interior point of the plaquette $p^i$ and $\epsilon^{ab}$ is the
Levi-Civita tensor. The basic observation is that the holonomy
$U_{p^i}\in SU(2)$ around the plaquette $p^i$ can be written as
\[U_{p^i}[A]=\mathbbm{1}+ \epsilon^2 F_{p^i}(A)+{\cal O}(\epsilon^2)\]
which implies \be F[N]=\int\limits_{U} {\rm Tr}[ N
{F}(A)]=\lim_{\epsilon\rightarrow 0}\ \sum_{p^i} {\rm
Tr}[N_{p^i}U_{p^i}[A]]\;,\ee where the $Tr$ in the r.h.s. is taken
in the fundamental representation. Notice that the explicit
dependence on the regulator $\epsilon$ has dropped out of the sum
on the r.h.s., a sign that we should be able to remove the
regulator upon quantization. With all this it is natural to write
the following formal expression for the generalized projection
operator: \ba \label{twenty} \ssp & = & \lim_{\epsilon\rightarrow
0} \ \ <\prod_{p^i} \ \int \ dN_{p^i} \ {\rm exp}(i {\rm Tr}[
N_{p^i} \hat
  {U}_{p^i}]) s, \; s^{\prime}>\\  \nonumber
& = & \lim_{\epsilon\rightarrow 0} \  \ <\prod_{p^i} \
{\delta}(U_{p^i})s, \; s^{\prime}>,\label{P3} \ea where the last
equality follows from direct integration over $N_{p^i}$ at the
classical level; $\delta(U)$ being the distribution such that
$\int dg\ f(g) \delta(g)=f(\mathbbm{1})$ for $f\in {\cal
L}^2(SU(2))$. We can write $\delta(U)$ as a sum over unitary
irreducible representations of $SU(2)$, namely \be\label{carac}
{\delta}({U_{p^i}})=\sum_j \Delta_j \ {{\chi}}_j(U_{p^i}), \ee
where $\Delta_j=2j+1$ is the dimension of the $j$-representation
and, at the classical level, ${\chi}_j(U)$ is the character or
trace of the $j$-representation matrix of $U\in SU(2)$. In
contrast to the formal example in four dimensions, mentioned in
the introduction, the previous expansion has more chances to have
a precise meaning in the quantum theory as each term in the sum
can be promoted to a well defined self-adjoint operator in $\Hk$:
it corresponds to the Wilson loop operator in the
$j$-representation around plaquettes corresponding to the
quantization introduced in (\ref{ggcc}).

The object ${\delta}{({U_{p^i}})}$ is clearly not well defined as
an operator in $\Hk$ as for any $\psi \in \Hk$ the state $
{\delta}{({U_{p^i}})}\psi$ is non-normalizable. This fact might
seem as a problem  for the definition of the generalized
projection operator $P$. This is however a standard feature which
appears in dealing with group averaging techniques when the
constraints generate non compact orbits (as clearly is the case
with the curvature constraint). Solutions of the curvature
constraint lie outside $\Hk$. Thus $\Hp$ is not a subspace of
$\Hk$. Physical states correspond to `distributional states' in
$Cyl^*$, the dual of the dense set $Cyl\subset\Hk$. Through the
inner product we can identify elements $|s>$ of a spin network
basis in $Cyl$ to dual spin networks $<s| \in Cyl^*$. The
generalized projector operator $\PP:Cyl \rightarrow Cyl^*$ takes
any element $s\in Cyl$ and sends it to a linear form $P s \in
Cyl^*$; we denote the action of this linear form on any element
$s^{\prime}\in Cyl$ by  $Ps(s^{\prime}):=<Ps,s^{\prime}>$. The
physical inner product is defined as $<a,b>_{p}:=<Pa, b>$. As a
result $\Hp$ can be viewed as the set of equivalence classes
elements of $Cyl^*$ of the form $<Ps|$ where $<Ps|\sim
<Ps^{\prime}|$ if they are equal as linear forms. The fact that
the limit (\ref{final1}) exists for $g\ge2$ for any
$s,s^{\prime}\in Cyl$ shows that the map $P$ is well defined.

Instead of working all the time with the explicit expression for
the physical inner product $(\ref{final1})$ it will be often
convenient to use the more compact expression  of $P$ in terms of
product of delta distributions as in (\ref{twenty}). \vspace{.5cm}

We conclude this section with some remarks:

\vspace{.2cm} \noindent {\em Remark 1:} The argument
of the limit in (\ref{final1}) satisfies the following bound
 \be \label{ecu} |<s,s^{\prime}>_p|=\left|\sum_{j_{\va p^i}}
 (2j_{\vani p^i}+1)\
 \mu_{AL}\left( \prod_{p^i} \chi_{j_{\va p^i}}({U}_{p^i}[A])\ \overline{ s[A]} s^{\prime}[A]\right)\right| \le  C \sum_{j}
 (2j+1)^{2-2g},\  \ee
where $C$ is a real positive
constant.\footnote{Let us define a regularization of the $SU(2)$
delta distribution as follows. Take a one parameter family of
(suficiently smooth) functions $\delta_{\Delta}(g)$ such that
$\delta_{\Delta}(g)\ge 0$ for all $g\in SU(2)$ and $\Delta\in\R^+$
and $\delta(g)=\lim_{\Delta\rightarrow 0}\delta_{\Delta}(g)$ in
the sense of distributions. We have the following identity
\be\label{yyy}<\prod_{p}  \sum_{j_{p}}(2j_p+1) \chi_{j_{p}}(U_{p})s,s'>\\
 =  \lim_{\Delta\rightarrow 0}<\prod_p
\delta_{\Delta}(U_p)s,s'>\;,\ee 
where we have
used two equivalent definitions of the delta function.
Due to the boundeness of the
elements of $Cyl$ we have
\[
R_1 \le {\rm Re}\left[\overline{ s[A]} s^{\prime}[A] \right] \le
R_2 \ \ \ {\rm and} \ \ \  I_1 \le {\rm Im}\left[\overline{ s[A]}
s^{\prime}[A] \right] \le I_2
\]
where $R_1,R_2, I_1, I_2\in \R$. Using the positivity of the regularization of
the delta distribution we can write \ba && \nonumber R_1 \sum_{j}
(2j+1)^{2-2g} = R_1 \lim_{\Delta\rightarrow 0} < \prod_p
\delta_{\Delta}(U_p);1> \le \\ \nonumber && \le \lim_{\Delta\rightarrow 0}
{\rm Re}< \prod_p \delta_{\Delta}(U_p) s; s^{\prime}> \le \\ \nonumber && \le
R_2 \lim_{\Delta\rightarrow 0} < \prod_p \delta_{\Delta}(U_p);1> = R_2
\sum_{j} (2j+1)^{2-2g}, \ea where the evaluation $\lim_{\Delta\rightarrow 0} <
\prod_p \delta_{\Delta}(U_p);1>=\sum_{j} (2j+1)^{2-2g}$ follows directly from
the definition of the AL-measure, see \cite{etera}. A similar bound holds for the
imaginary part from where one can deduce (\ref{ecu}) using the fact that \[
\sum_{j_{p}}(2j_p+1) <\prod_{p}   \chi_{j_{p}}(U_{p})s,s'>= <\prod_{j_{p}}
\sum_{j_{p}}(2j_p+1) \chi_{j_{p}}(U_{p})s,s'>,\] once the convergennce of the
r.h.s. is granted.}  
The convergence of the sum for $g\ge 2$ follows directly.
\vspace{.2cm} 

\noindent {\em Remark 2:} The case of the sphere
$g=0$ is easy to regularize. In this case (\ref{final1}) diverges
simply because of a redundancy in the product of delta
distributions in the notation of (\ref{twenty}). This is a
consequence of the discrete analog of the Bianchi identity. It is
easy to check that eliminating a single arbitrary plaquette
holonomy from the product in (\ref{final1}) makes $P$ well defined
and produces the correct (one dimensional) $\Hp$.

The case of the torus $g=1$ is more subtle; we will discuss it at
the end of the paper.

 \vspace{.2cm}
\noindent {\em Remark 3:} Clearly the order of the product of
$\delta$-distributions is not important, as the operators being
multiplied are all commuting. For this reason we will consider a
symmetrized version of the previous regularization that it is
trivially equivalent to (\ref{twenty}) in this case \footnote{ As
we shall shortly see, each of the $\delta$ distribution produces a
local transition in the spin foam representation. Different
orderings correspond to different `coordinate time' evolutions and
produce different consistent (histories) {\em spin foams}. Now all
these spin foams have the same amplitude in this theory but in a
general setting should represent independent contributions to the
path integral. }. Thus from now on we will consider the following
expression of $P$ :
\begin{equation}\label{PS}\ssp:= \ \lim_{\epsilon\rightarrow 0} \ \frac{1}{N^{\epsilon}_p!} \sum_{\sigma} \
\ < \prod_{p^{\sigma(i)}} \ {\delta}({U}_{p^{\sigma(i)}})s, \  s^{\prime}>,\end{equation}
where $N^{\epsilon}_p$ is the $\epsilon$-dependent number of plaquettes and $\sigma(\{i\})$ denotes a permutation of the
set of plaquette labels. Now we introduce some notation. We denote by $\Pe$
the argument of the limit above which we can think of as a truncated version
of $P$, namely:
\begin{equation}\label{Pe} \hat{\Pe}:=  \frac{1}{N^{\epsilon}_p!} \sum_{\sigma(\{i\})} \
\ <\prod_{p^{\sigma(i)}} \ {\delta}({U}_{p^{\sigma(i)}}) s \ s^{\prime}>.\end{equation}
We denote by $\Pes$ each of the possible orderings $\sigma$ of delta
distributions in the previous expression, explicitly:
\begin{equation}
\label{Pes} \hat{\Pes}:= <\prod_{p^{\sigma(i)}} \
{\delta({U}_{p^{\sigma(i)}})} s \ s^{\prime}>.
\end{equation}

\vspace{.2cm}

\noindent {\em Remark 4:} It is immediate  to see that
(\ref{final1}) satisfies hermitian condition \be\ssp=\sspb.\ee

\vspace{.2cm}

\noindent {\em Remark 5:} The positivity condition also follows
from the definition $\sspe\ \ge 0$.

\subsection*{3.2. Spin foam representation}

As in the case of the standard Feynman derivation of the path
integral representation of the evolution operator, the spin foam
representation of the generalized projection operator follows from
inserting resolutions of the identity in $\Hk$ given by \be
\mathbbm{1}=\sum \limits_{\Gamma \subset \Sigma, \{j\}_{\Gamma}}
|\Gamma,\{j\}><\Gamma,\{j\}| \label{e} \ee between consecutive
delta distributions in (\ref{PS}). In the previous expression the
sum is taken over all the elements of a chosen spin network basis
in $\Hk$. Because of the definition of the Ashtekar-Lewandowski
measure only those spin networks that differ by a single plaquette
loop contribute in following steps making the transition
amplitudes finite.

\begin{figure}[h!!!!!]
 \centerline{\hspace{0.5cm}\(
\begin{array}{ccc}
\includegraphics[height=1.9cm]{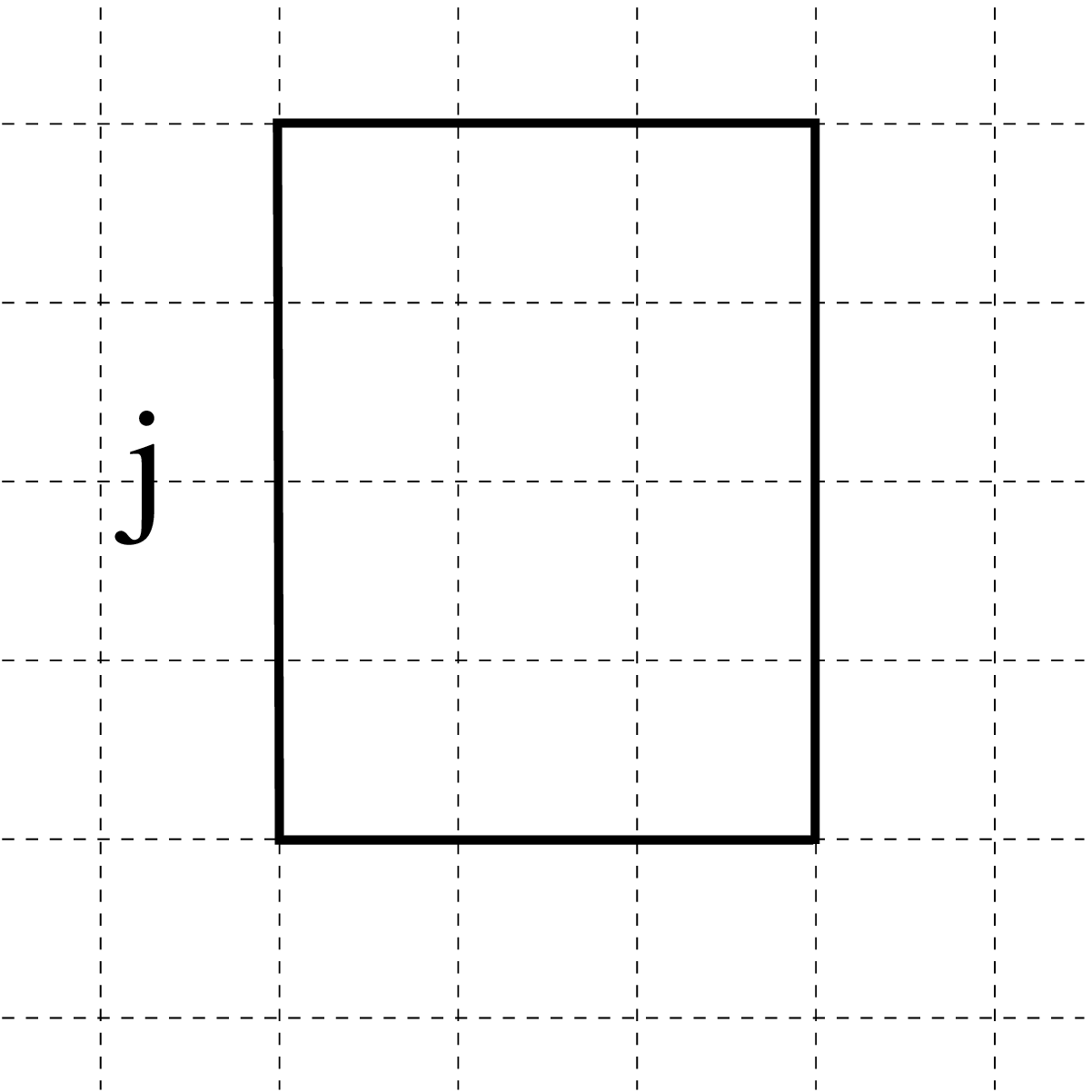} \\
\includegraphics[height=1.9cm]{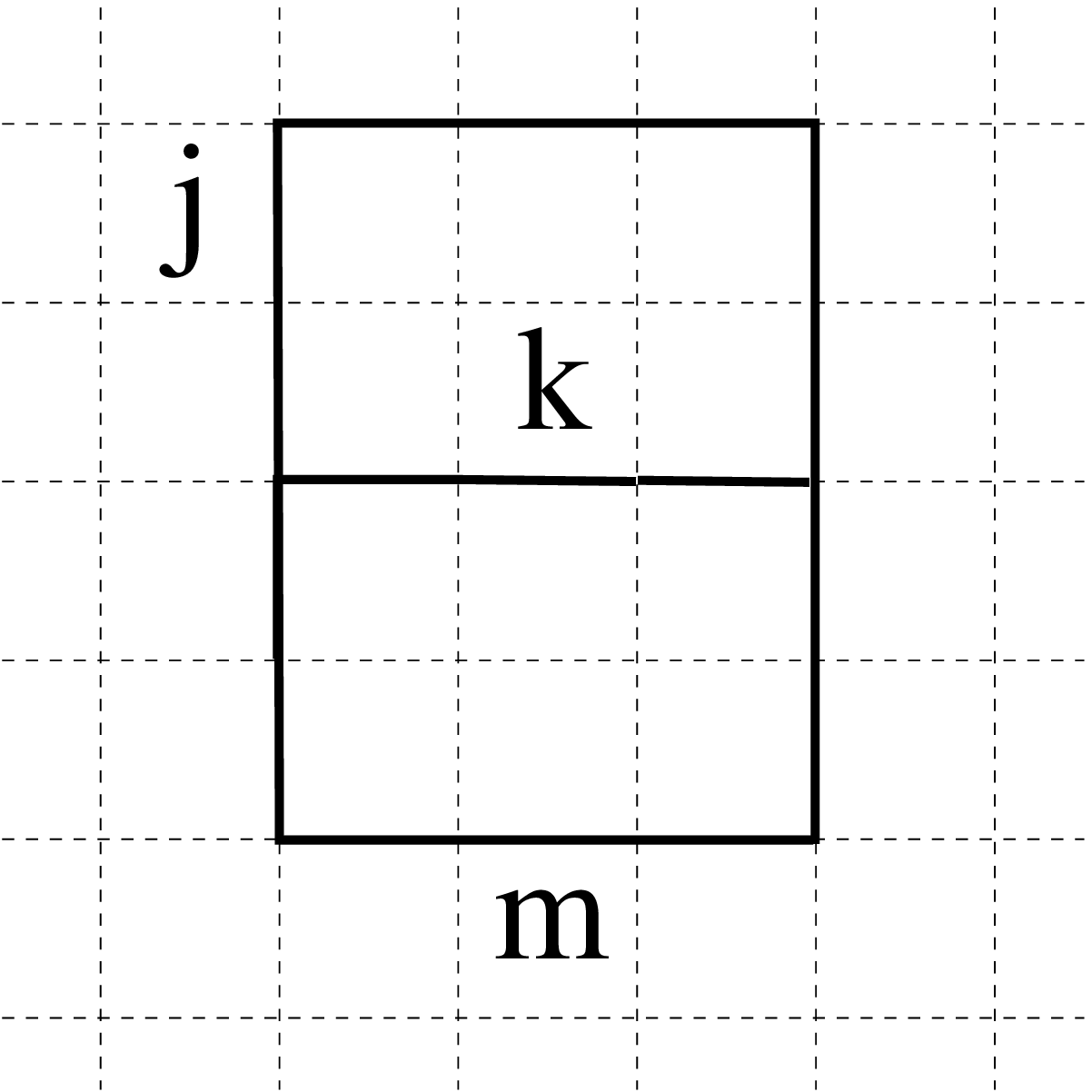}
\end{array}
\begin{array}{ccc}
\includegraphics[height=1.9cm]{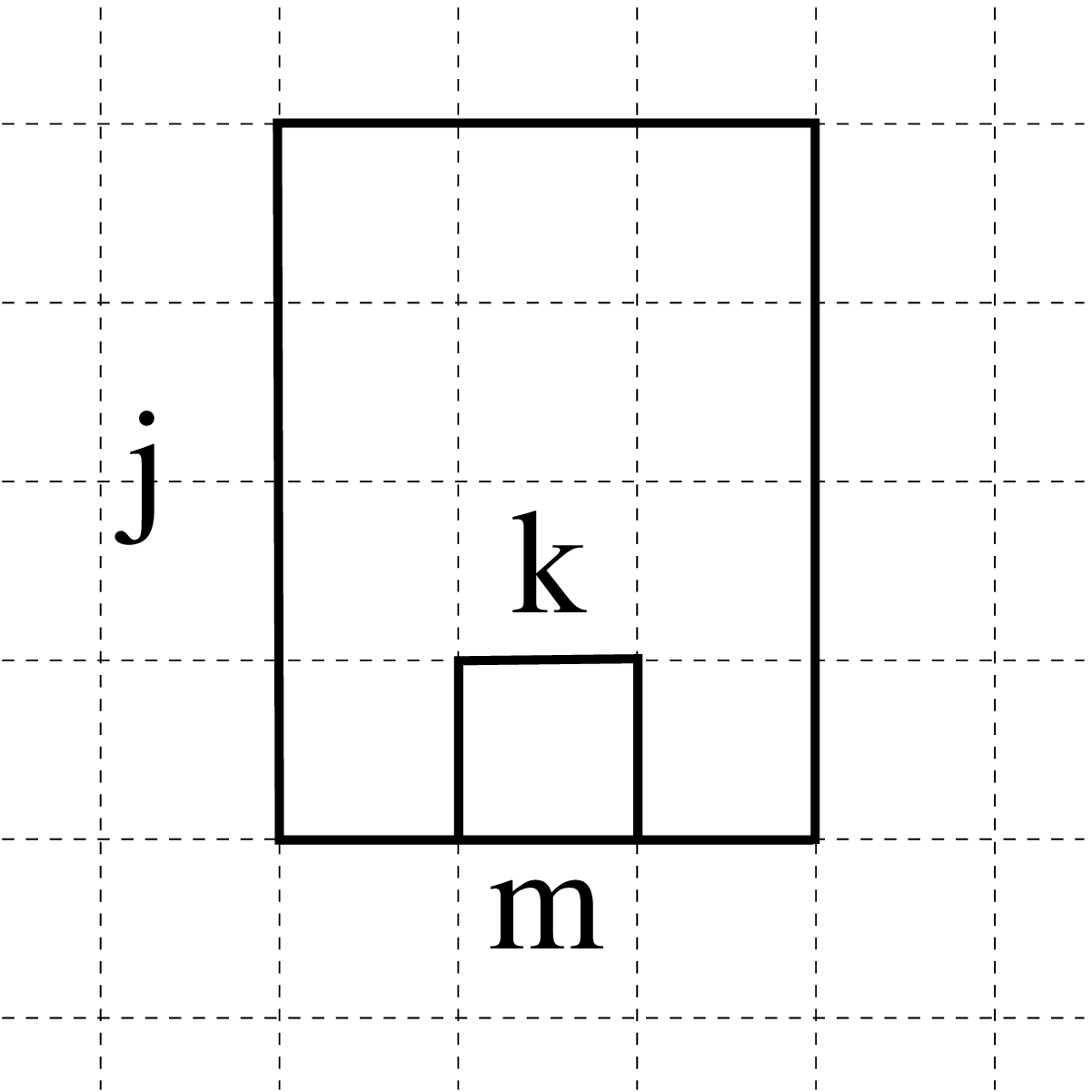} \\
\includegraphics[height=1.9cm]{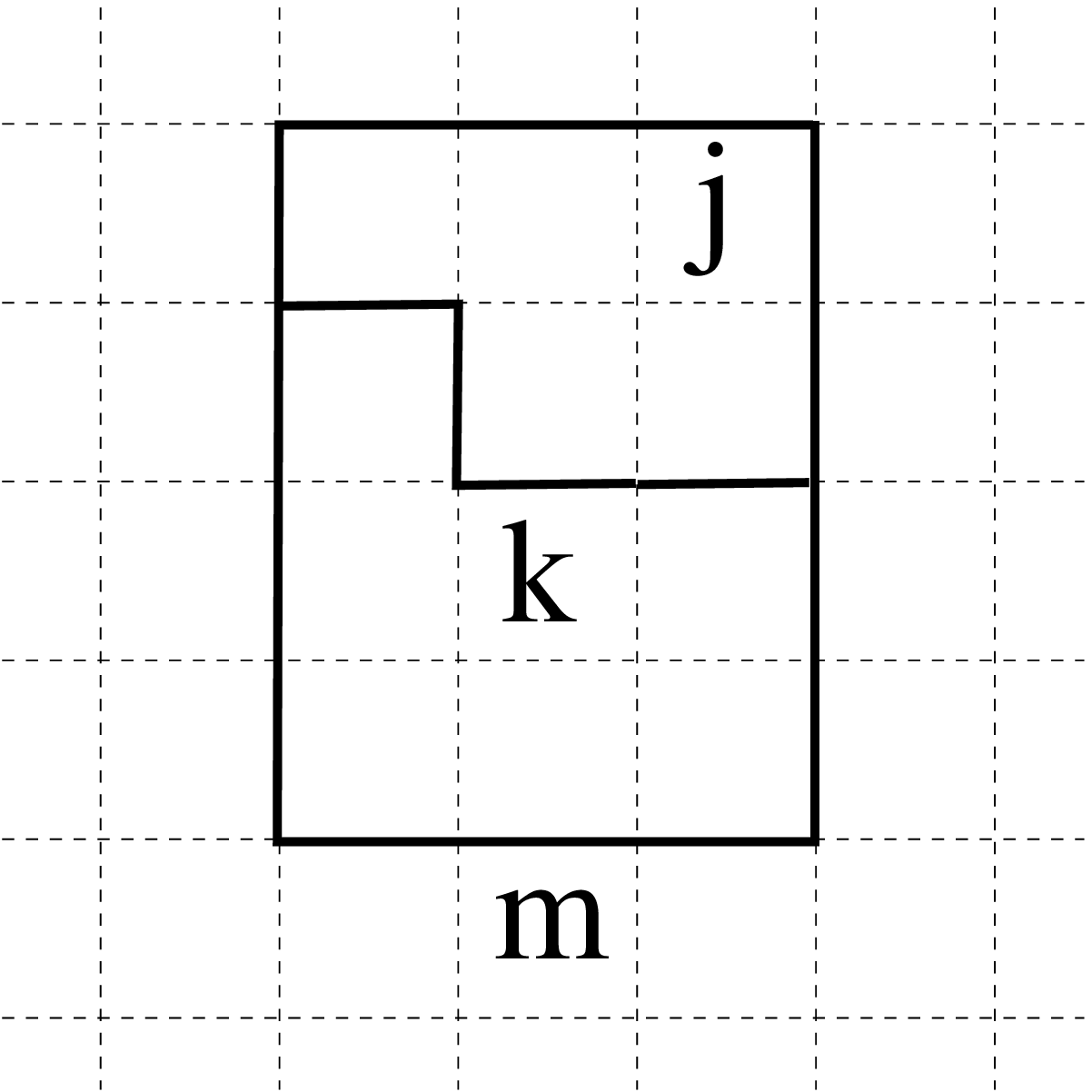}
\end{array}
\begin{array}{ccc}
\includegraphics[height=1.9cm]{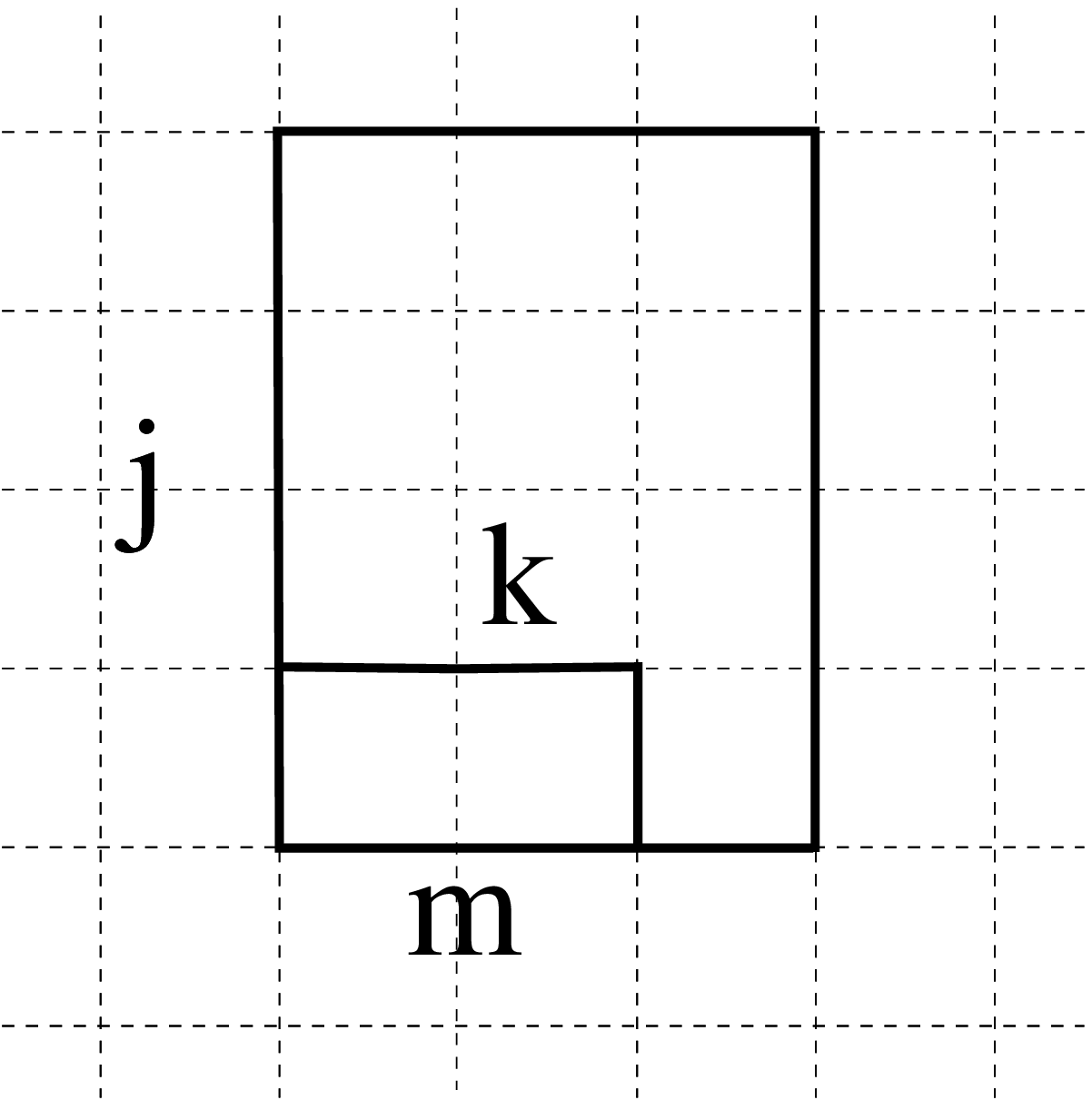} \\
\includegraphics[height=1.9cm]{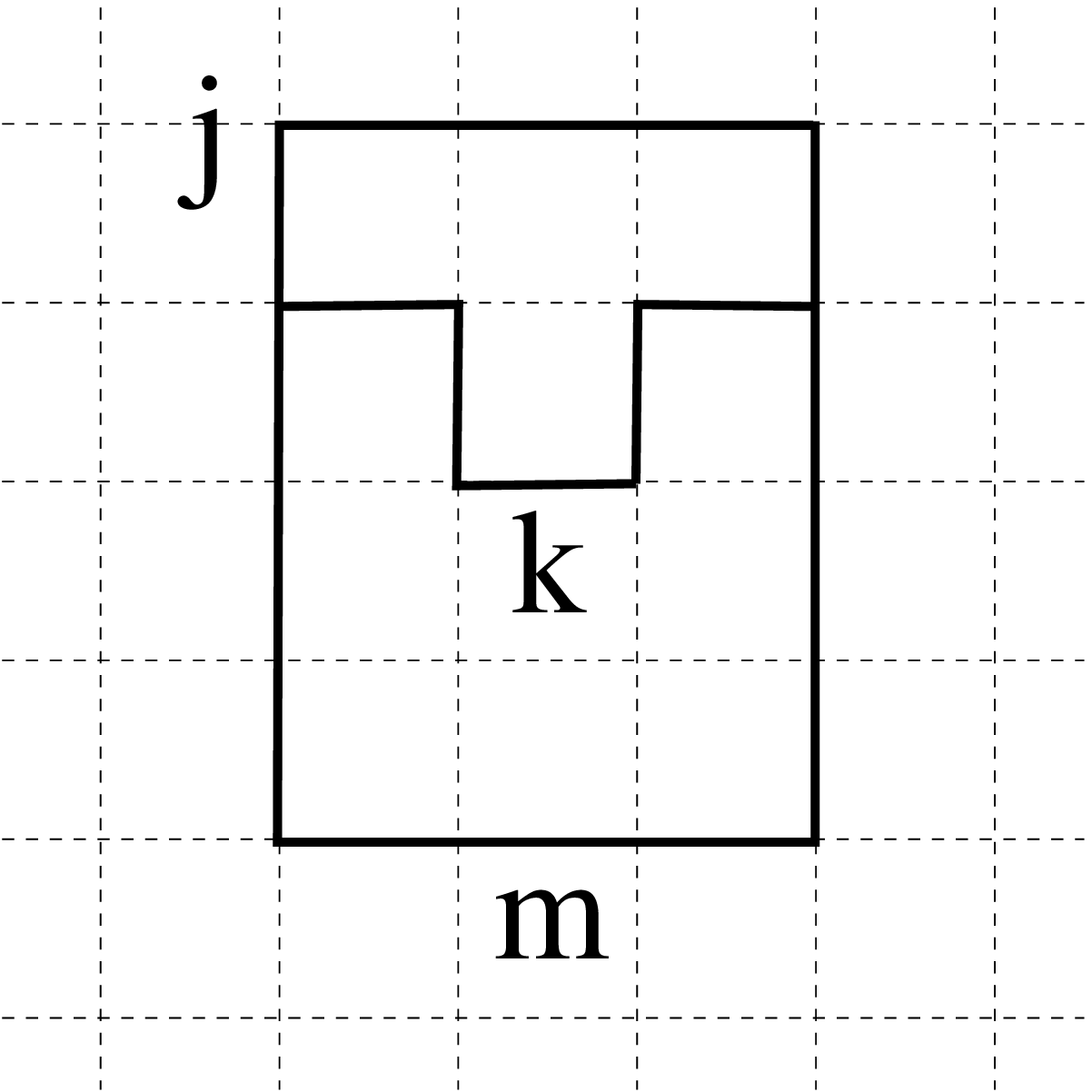}
\end{array}
\begin{array}{ccc}
\includegraphics[height=1.9cm]{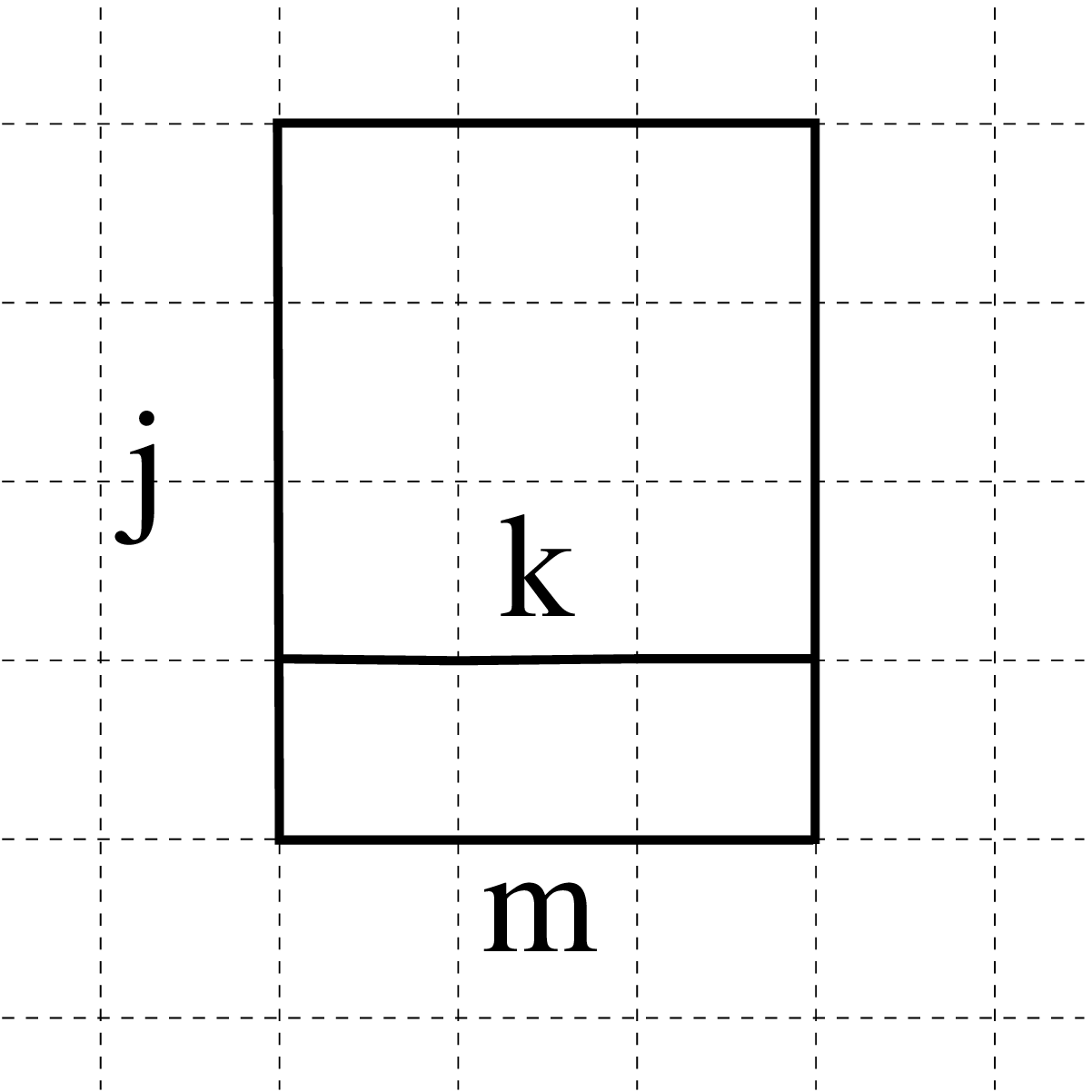} \\
\includegraphics[height=1.9cm]{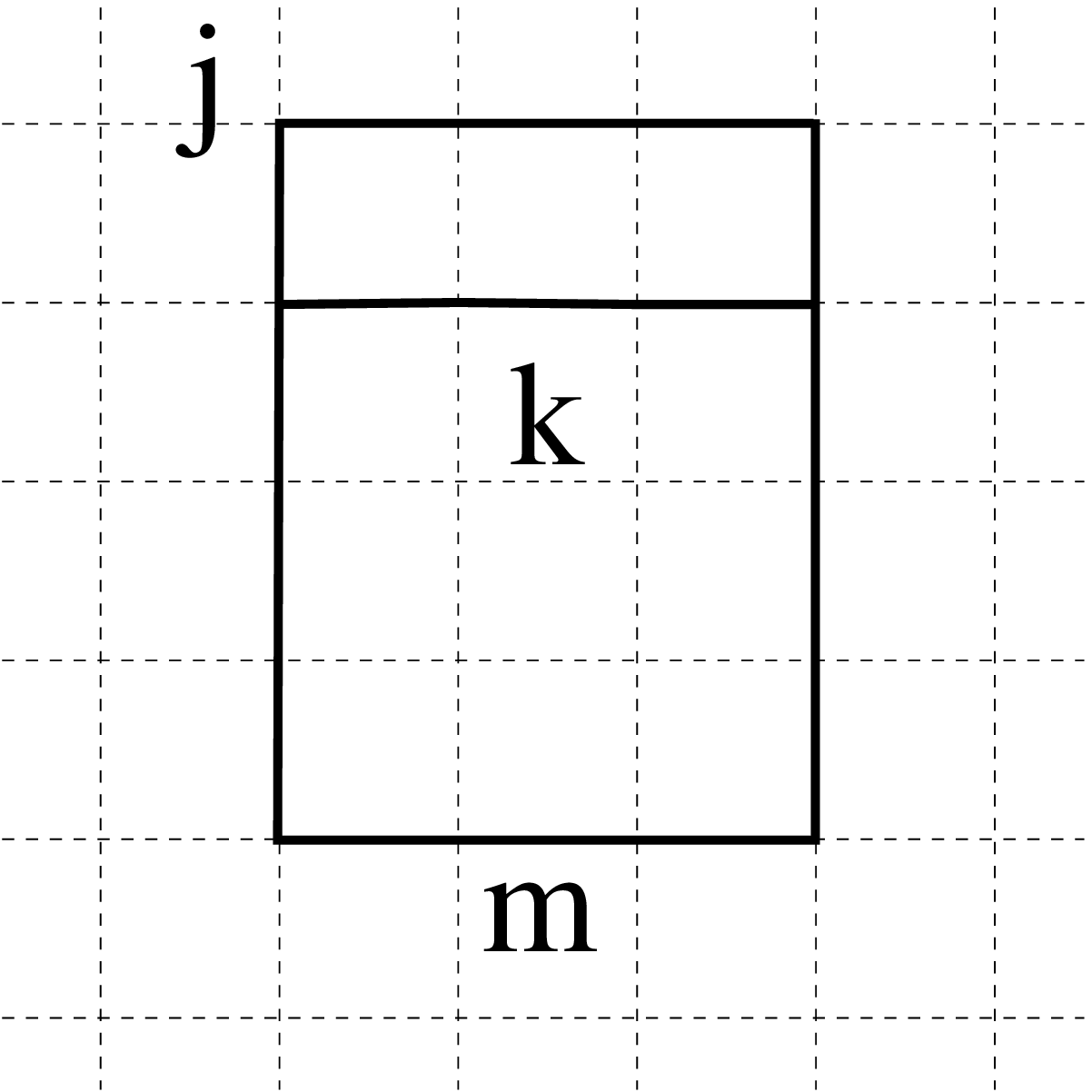}
\end{array}
\begin{array}{ccc}
\includegraphics[height=1.9cm]{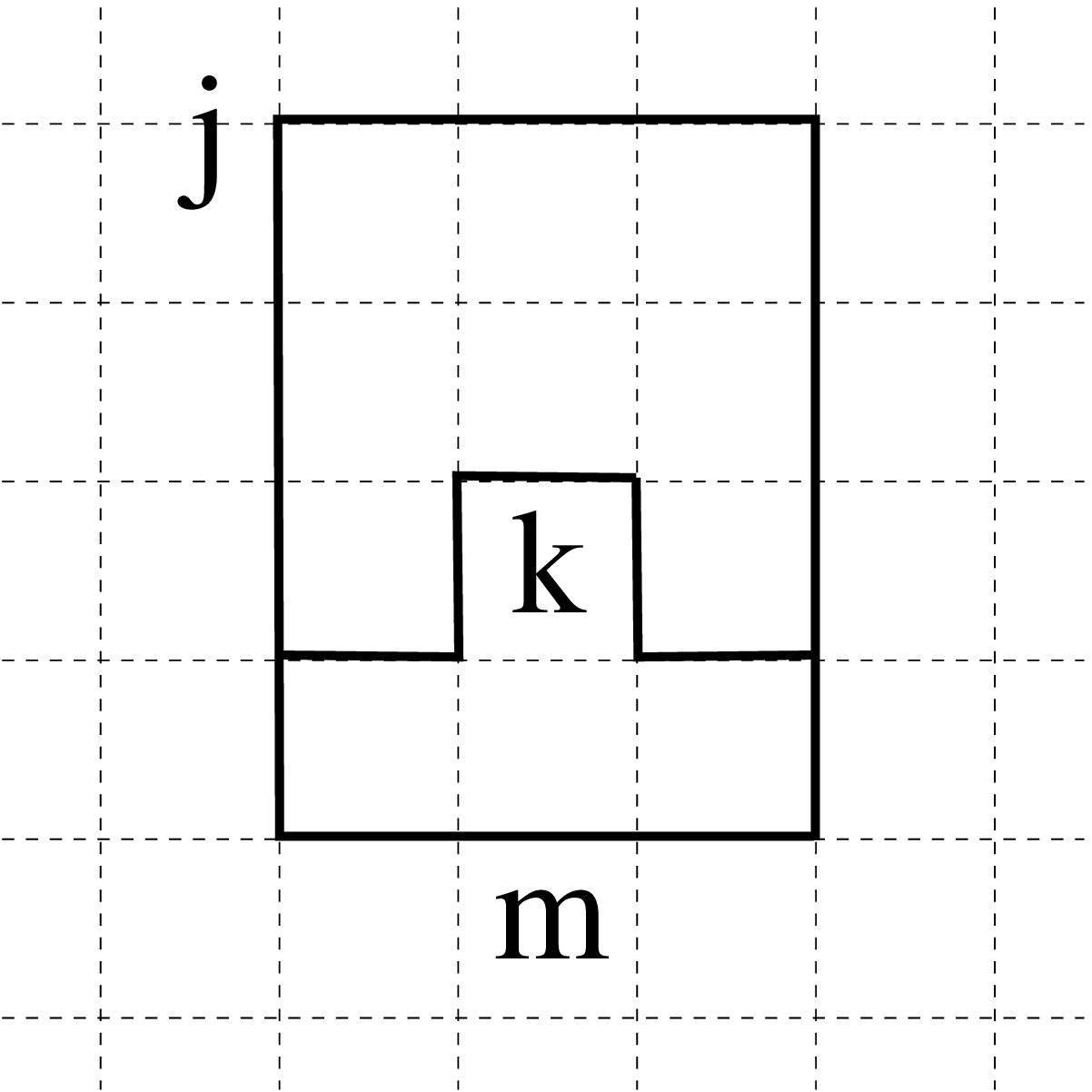} \\
\includegraphics[height=1.9cm]{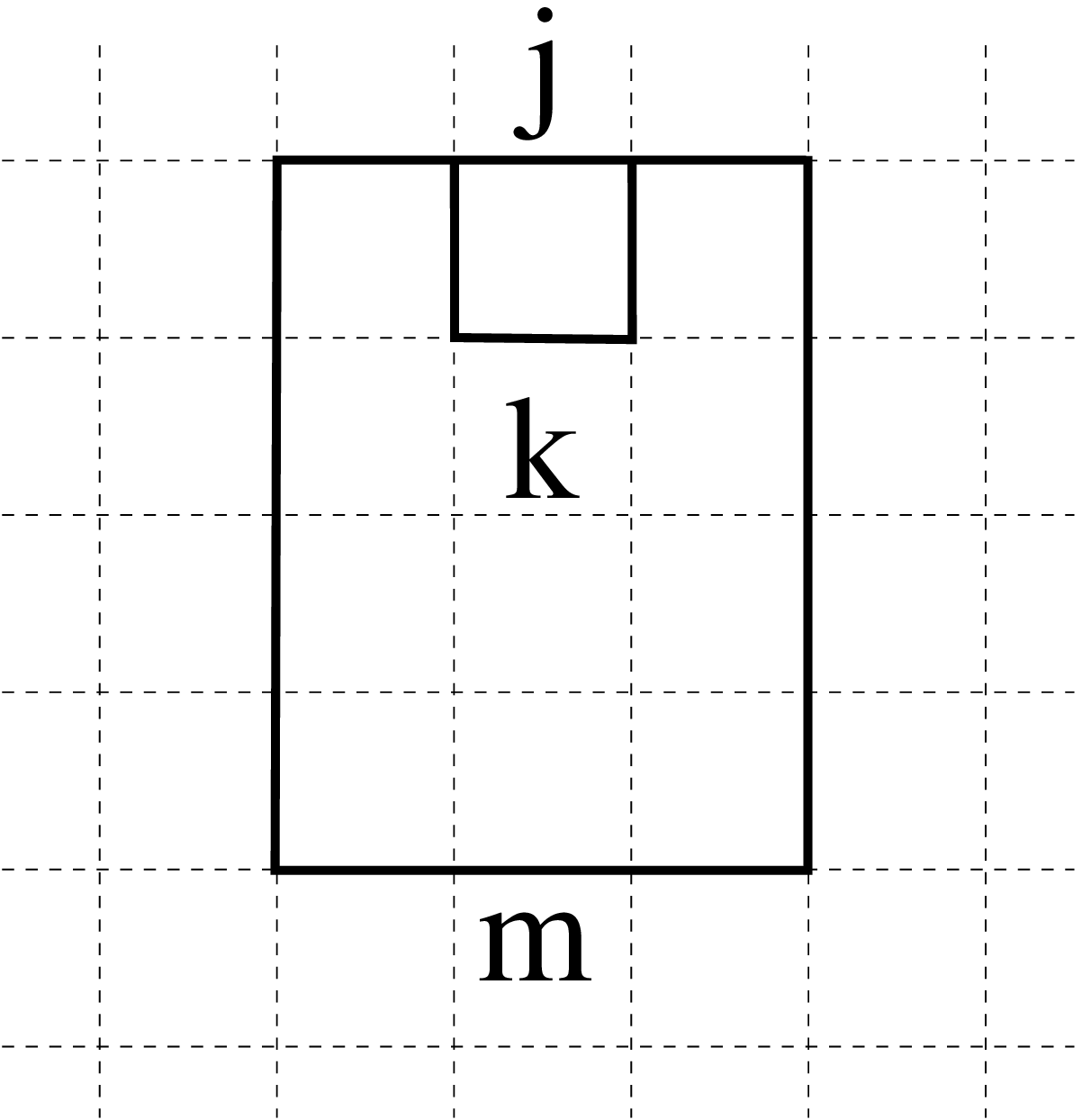}
\end{array}
\begin{array}{ccc}
\includegraphics[height=1.9cm]{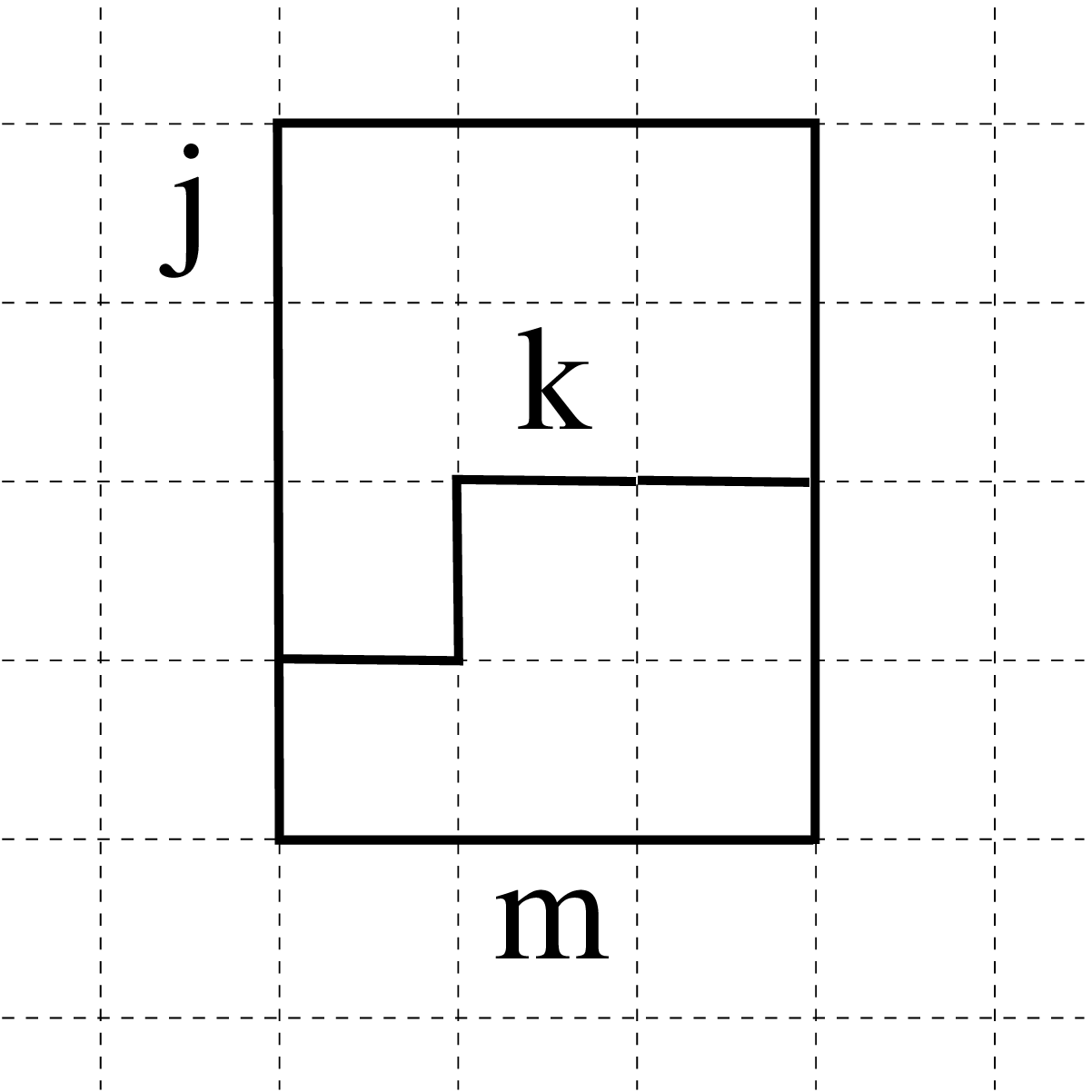} \\
\includegraphics[height=1.9cm]{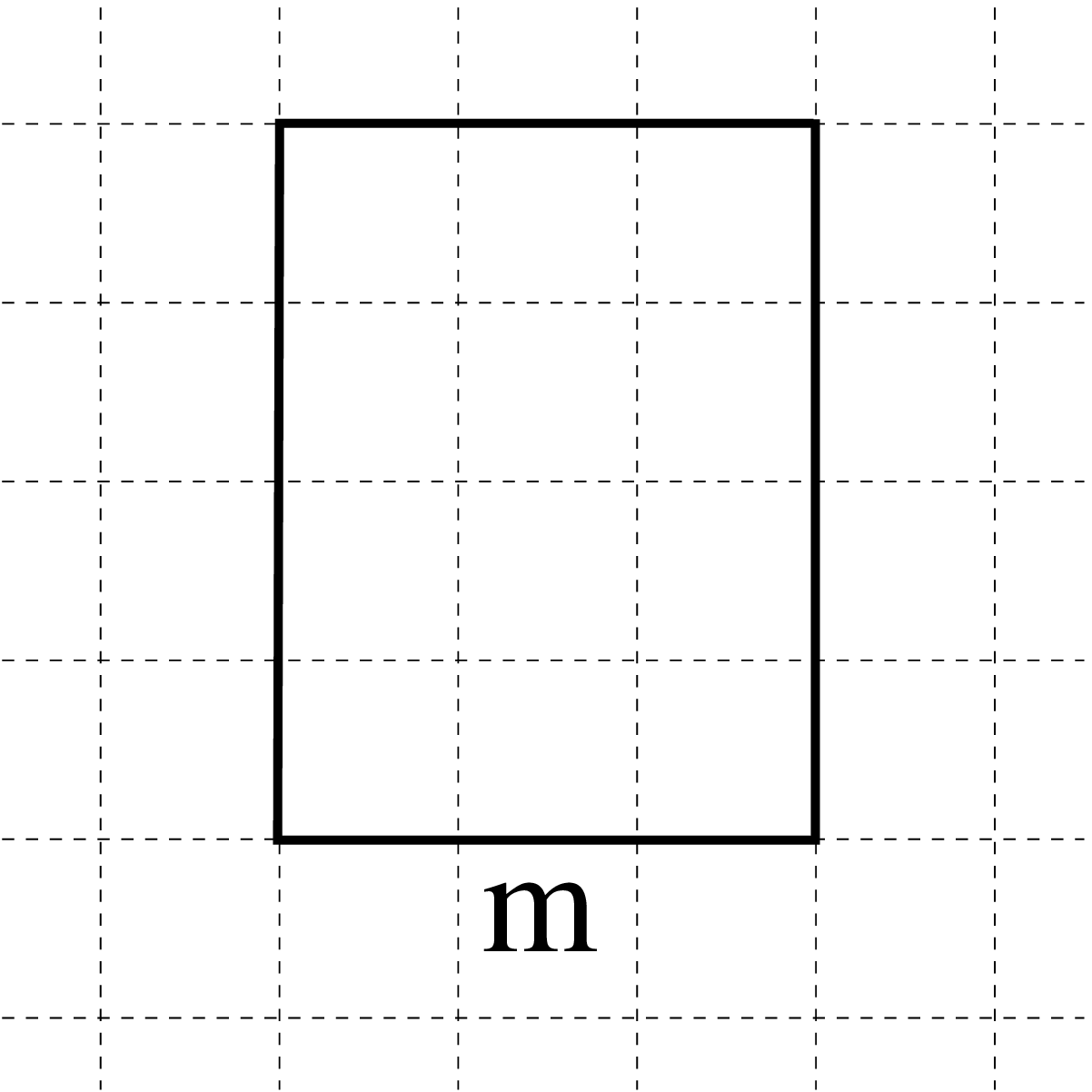}
\end{array}
\)}
\caption{\small A set of discrete transitions representing one of the contributing histories implied
by our regularization of the generalized projection $P$
in Equation (\ref{PS}); from left to right in two rows.}
\label{lupy}
\end{figure}

In order to visualize in which way the spin foam representation arises, let us
consider a simple example. For concreteness we assume that the value of the
regulator $\epsilon$ is fixed at some non zero value. Consider the regularized
matrix element $\Pes$ between the spin networks $s$ and $s^{\prime}$ based on
the loop $\ell  \in  \Sigma$ and labeled by the
representations $j$ and $m$ respectively (see Figure \ref{lupy}).
At this finite regulator $\epsilon$ we have $9$ plaquettes covering the
interior of the loop. Taking the definition of $\Pes$ and
inserting resolutions of identities between the plaquette delta distributions
it is easy to show that from the large set of graphs in $\Sigma$
only a finite number of intermediate graphs survive in
the computation of $\Pes$.

Let us study the first step in the sequence of transitions illustrated in
Figure \ref{lupy}. We consider the ordering of plaquettes such that the first
delta distribution in the regularization of $P$ (one term in the sum over
orderings of equation (\ref{PS})) is evaluated on the lower
central plaquette with respect to the loop $\ell$. The first delta function
acts on the loop $\ell$ as
\ba
&& \nonumber
{\delta}(U_p)\begin{array}{c}
\includegraphics[width=2.5cm]{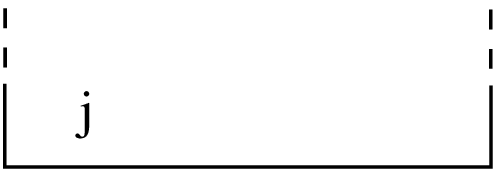}
\end{array}
= \\ &&  \centerline{\hspace{0.5cm} \(
\sum \limits_s \Delta_s \begin{array}{c}
\includegraphics[width=2.5cm]{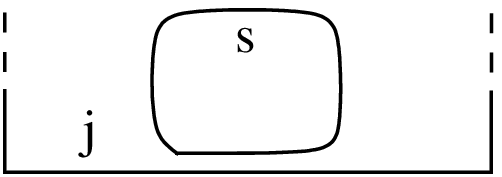}
\end{array}=\sum\limits_{k,s} \Delta_k N_{j,m,k} \delta_{k,s}
\begin{array}{c}
\includegraphics[width=2.5cm]{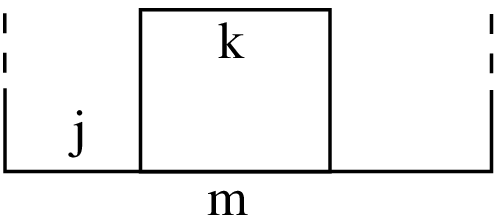}
\end{array}
\) }
\label{pito}
\ea
where on the right hand side we have expanded the result of the
action of the Wilson loop operator in the representation $s$ on our initial
state in terms of spin network states. The coefficient of the expansion is
determined by the AL measure, namely
\ba
\centerline{\hspace{0.5cm} \(
 N_{j,m,k} \delta_{k,s}
= \begin{array}{c}
\includegraphics[height=1.5cm]{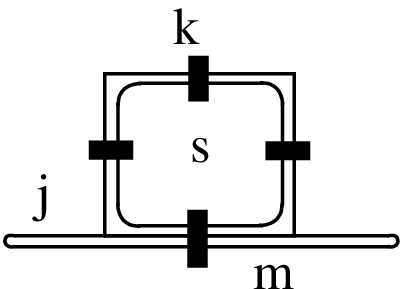}
\end{array}
,\) } \ea where we are using the graphical notation described in
the appendix (the Haar measure integration is denoted by a dark
box overlapping the different representation lines (see
(\ref{lulingui}))). It is a simple exercise to evaluate the
previous diagram resulting in $ N_{j,m,k}=1$ if $j,k,m$ are
compatible spins and $ N_{j,m,k}=0$ otherwise. Equation
(\ref{pito}) implies that when inserting the resolution of the
identity (\ref{e}) between the first and second delta
distributions the corresponding term in $\Pes$ only the
intermediate states on the right hand side of (\ref{pito}) will
survive. The first delta distribution acts on the initial state by
attaching a new infinitesimal loop. The following deltas would
have a similar effect creating new infinitesimal loops associated
to each corresponding plaquettes. In this way each term in $\Pe$
`explores' the set of intermediate spin network states based only
on those graphs that are contained in the regulating cellular
decomposition $\Seg$. In this set of transitions $SU(2)$ gauge
invariance is preserved which is manifested as spin compatibility
conditions. Finally, when we contract with the final state
$s^{\prime}$ only the sequence of spin network states which
represent consistent histories remain. The consistency condition
is precisely given by the requirement that the set of transitions
produces a spin foam \cite{baez7}.

An important fact is that these discrete spin foams do not contain bubbles.
Once a new loop labeled by a non-trivial spin is created it cannot be
annihilated. The Wilson loops operators in the series that define the delta
distributions are self adjoint operators and in this way they can both create
or annihilate infinitesimal loops. However, they can act only in one direction
when sandwiched between fixed boundary states. In order to produce a spin foam
with a bubble we would need to act twice with the same Wilson loop which
amounts to acting twice with the delta distribution; a clearly ill defined operation. In fact it would lead
to a divergence proportional to $\delta(\mathbbm{1})=\sum_s (2s+1)^2$. This is
precisely the kind of bubble divergences obtained in the Ponzano-Regge model!
No such divergences are present in our definition of $P$ as each plaquette
delta distribution acts only once (\ref{PS}).

\begin{figure}[h!!!!!]
\centerline{\hspace{0.5cm} \(\begin{array}{c}
\includegraphics[height=5cm]{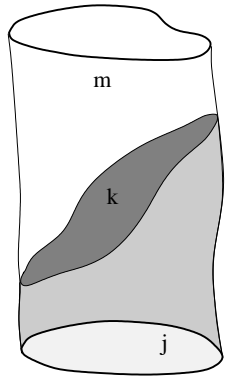}
\end{array}\) }
\caption{\small Spin foam representation of the transition between to loop states. Our regularization of the
generalized projection operator $P$ produces (in the path integral representation) a continuous transition
between embedded spin networks. Here we illustrate the result at three different slicing.}
\label{lupos}
\end{figure}

Therefore, the AL measure selects only those intermediate spin networks which provide a sequence of
transitions between the `initial' and `final' state that can be represented by
a spin foam with no bubble. In the case at hand and considering a particular arbitrary
ordering of the plaquette delta distributions one of such sequence is
illustrated in Figure \ref{lupy}. There is one such sequence for each value of
$k$ satisfying the spin compatibility conditions with $j$ and $m$ and its
amplitude is given by $\Delta_k$ as we will explicitly shown. Notice that this
is precisely the value of such spin foam in the Ponzano-Regge model.
In the limit $\epsilon\rightarrow 0$ the discrete sequences contain more and
more intermediate states and  tend to a
continuous history or continuous spin foam representation. One of these
possible histories is depicted in Figure \ref{lupos}.

Another example representing four valent vertex transition is represented in
Figure \ref{vani}. The continuous spin foam picture illustrated in Figure
\ref{pilin} is obtained when the regulator is removed in the limit
$\epsilon\rightarrow 0$.

We will show in the sequel that in the limit when the regulator is removed
we recover a spin foam representation of the matrix elements of the projection
operator and hence the physical Hilbert space $\Hp$ that is independent of any
auxiliary structure.

\begin{figure}[h!!!!!]
 \centerline{\hspace{0.5cm}\(
\begin{array}{ccc}
\includegraphics[height=2.5cm]{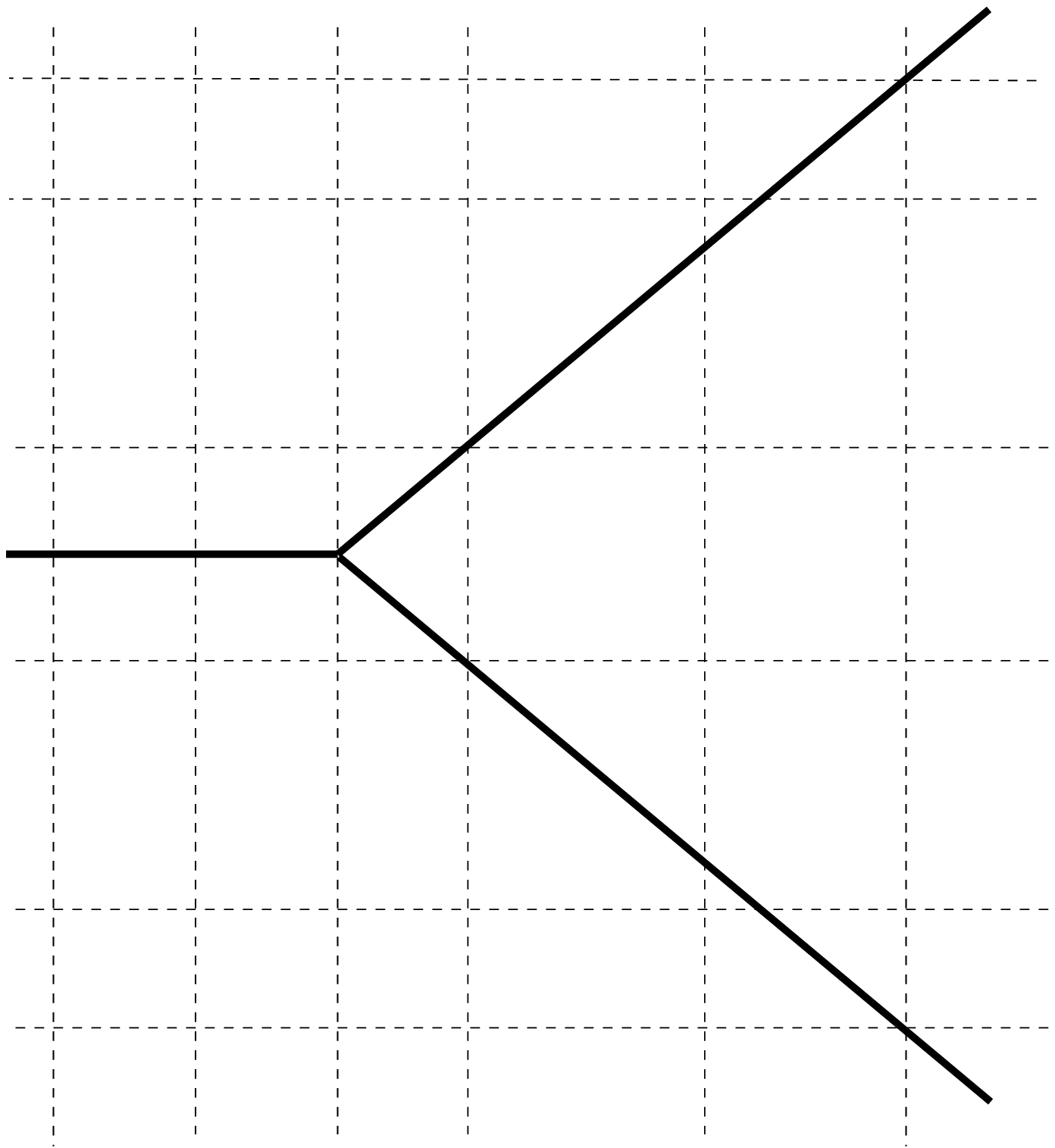} \\
\includegraphics[height=2.5cm]{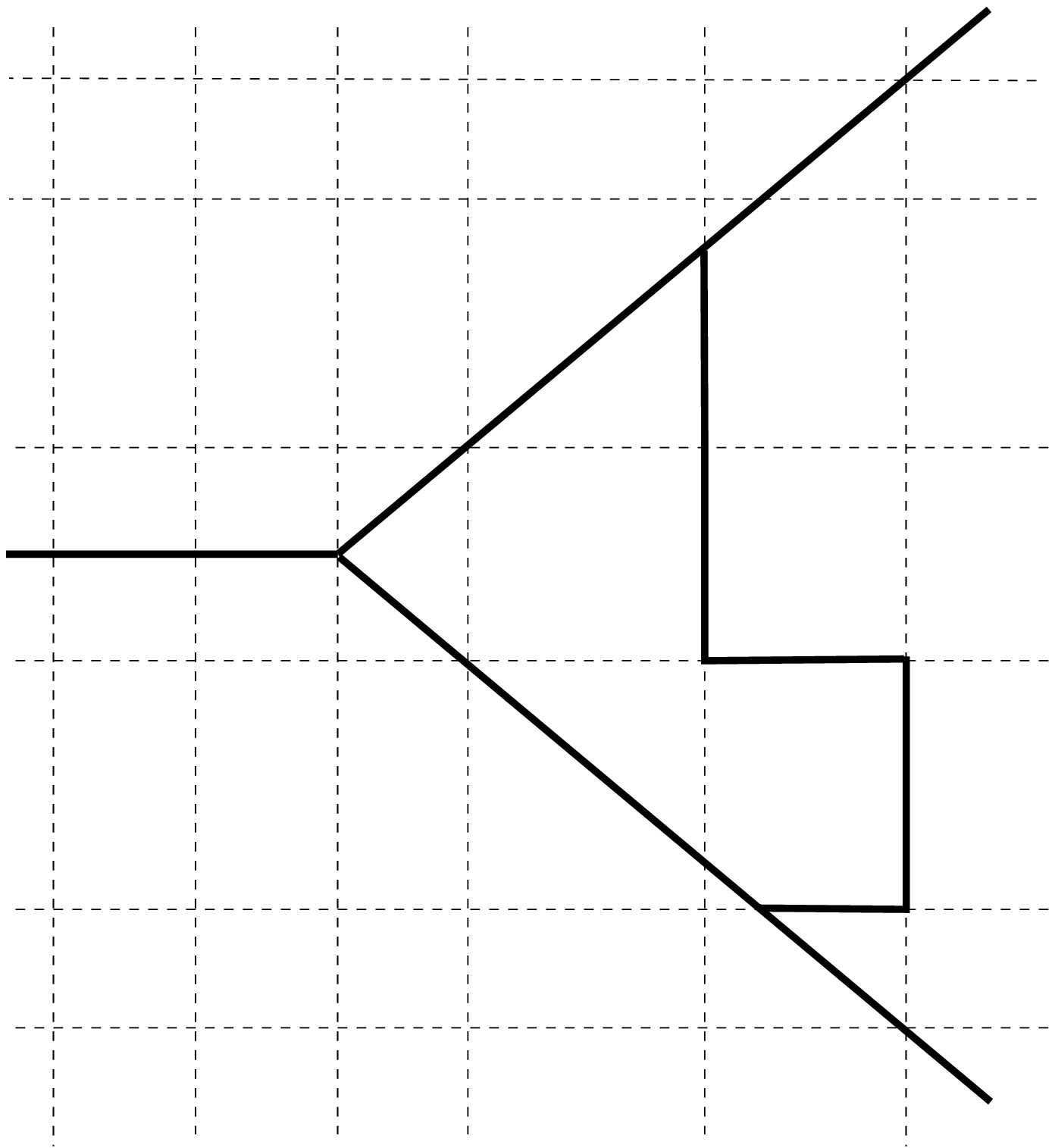}
\end{array}
\begin{array}{c}
\includegraphics[height=2.5cm]{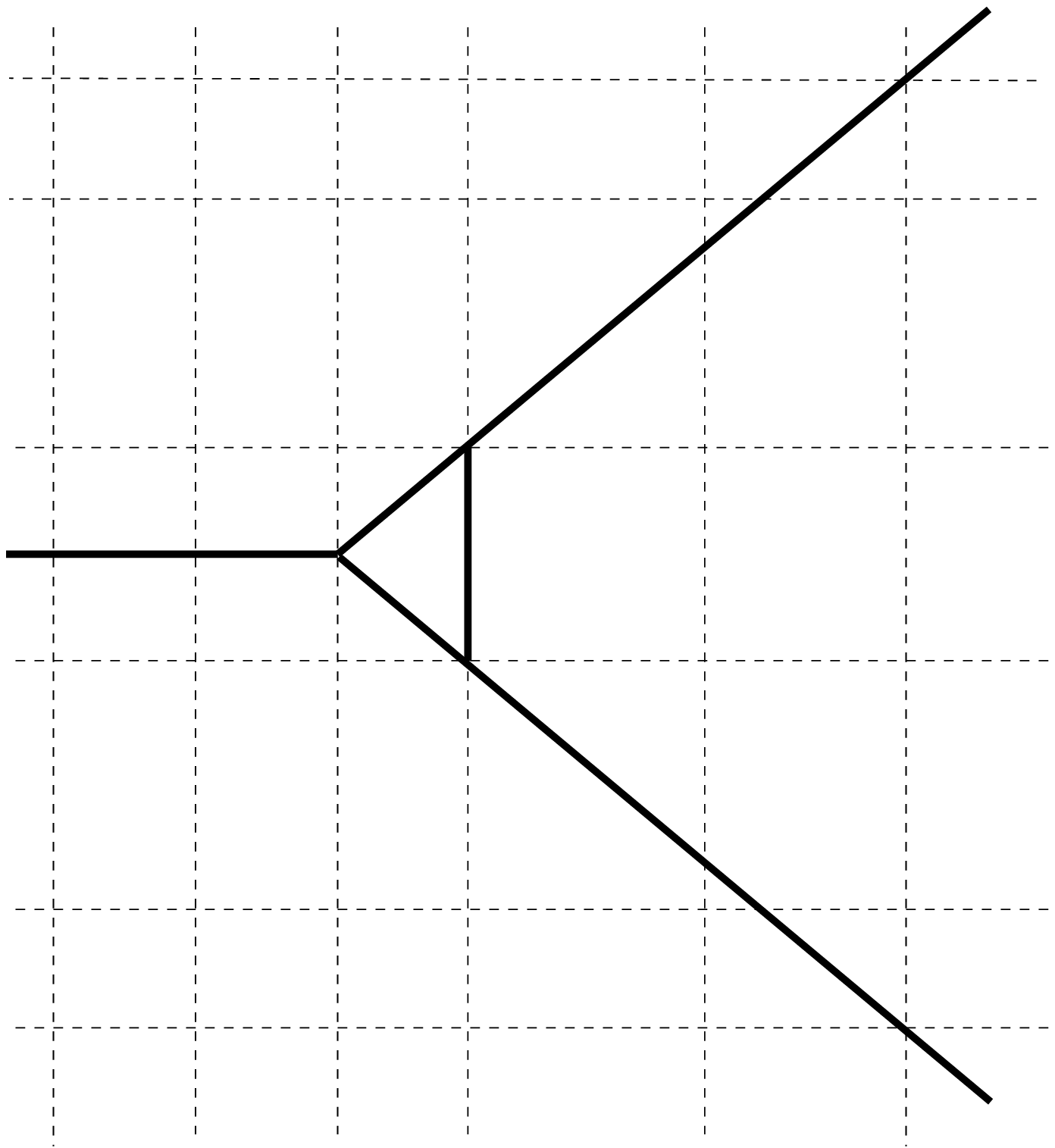}\\
\includegraphics[height=2.5cm]{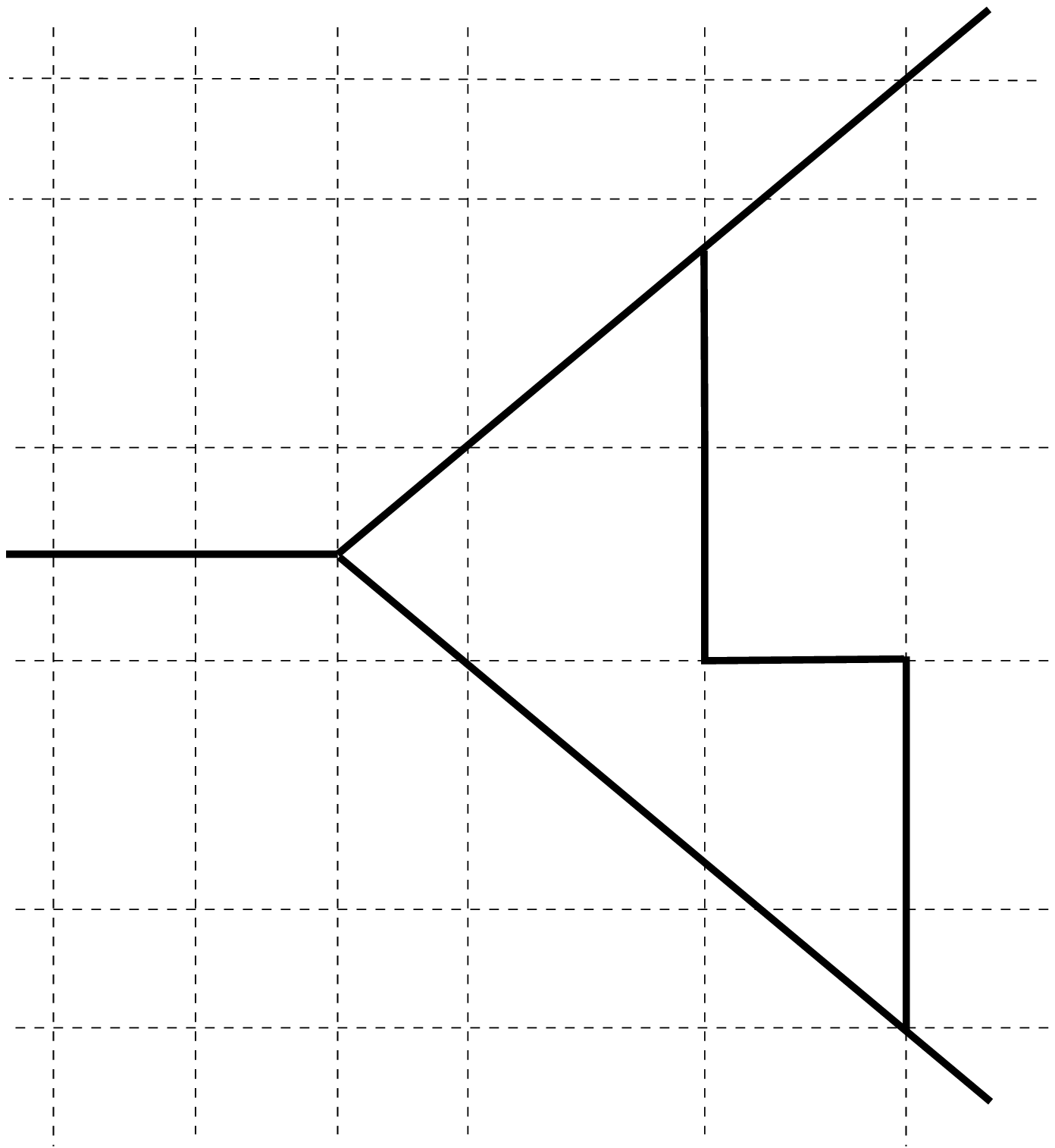}
\end{array}
\begin{array}{c}
\includegraphics[height=2.5cm]{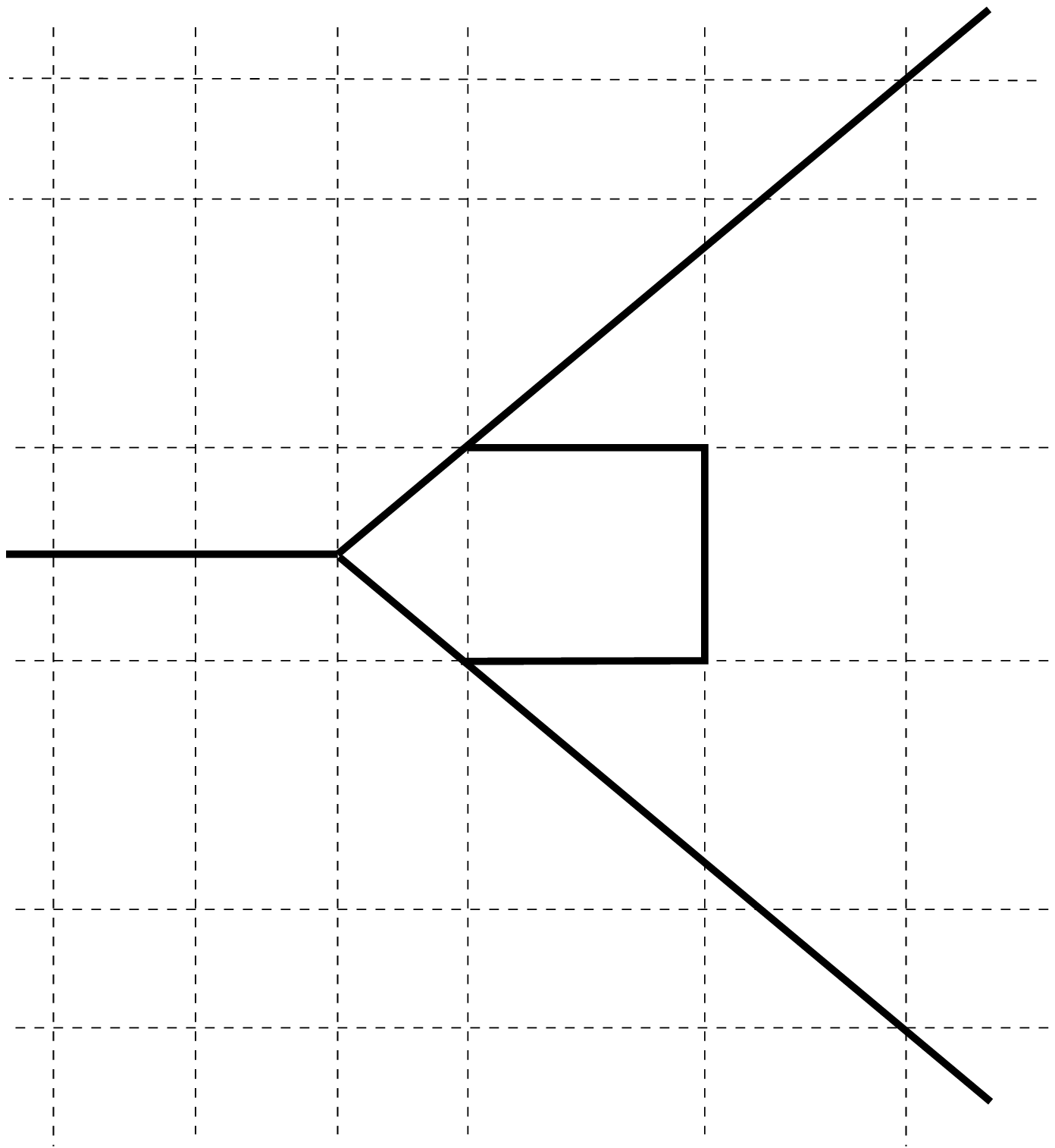}\\
\includegraphics[height=2.5cm]{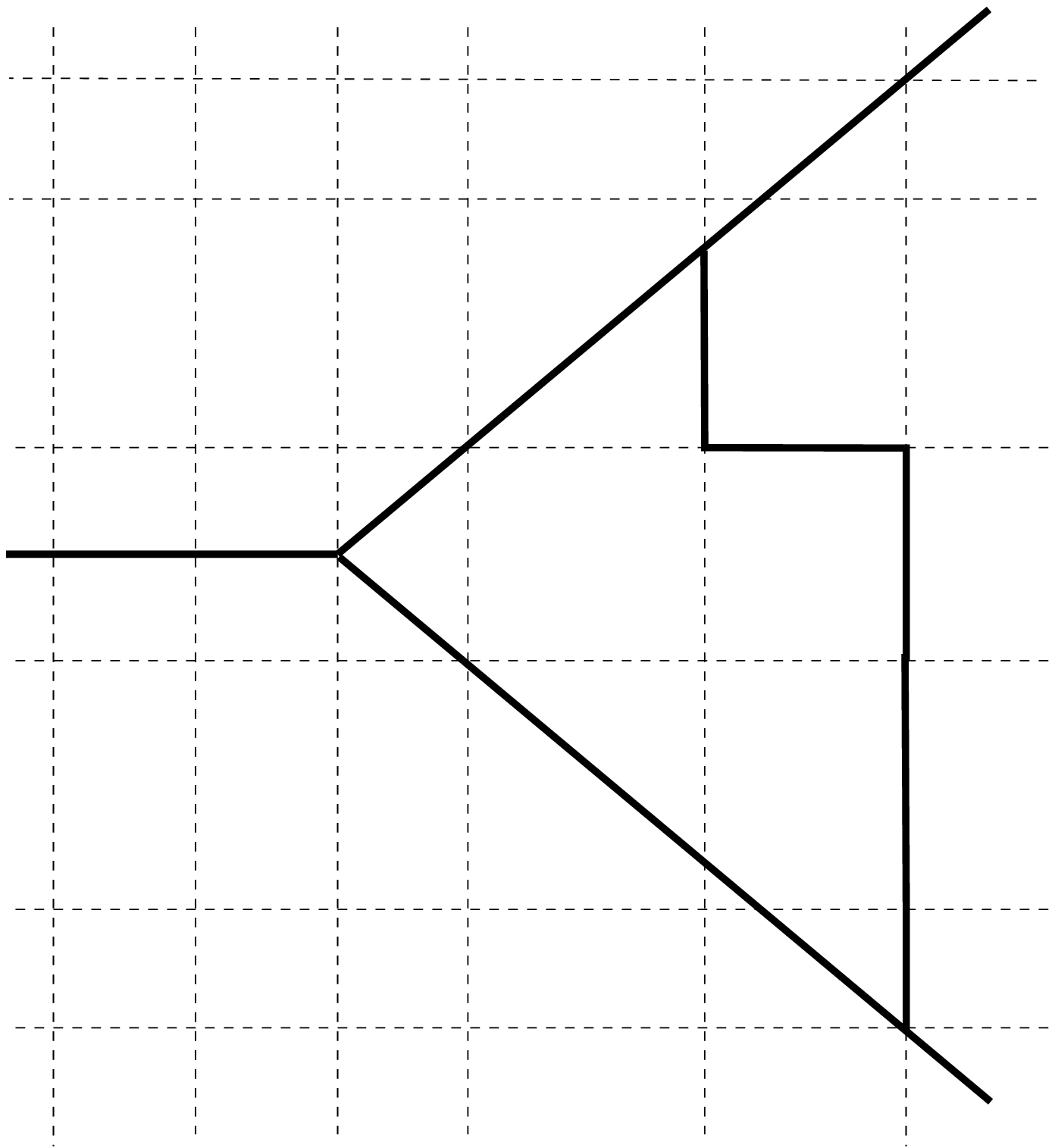}
\end{array}
\begin{array}{c}
\includegraphics[height=2.5cm]{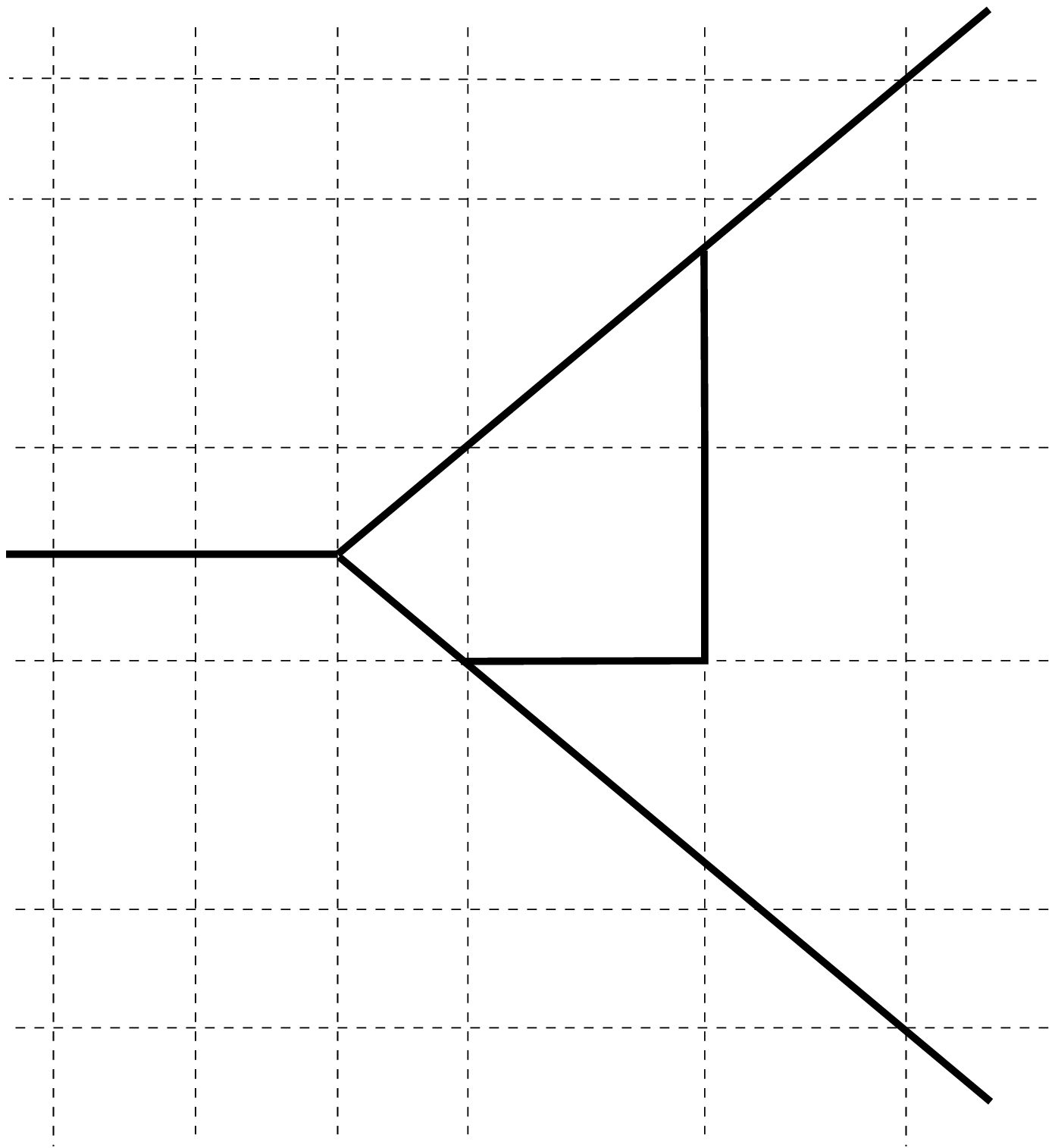}\\
\includegraphics[height=2.5cm]{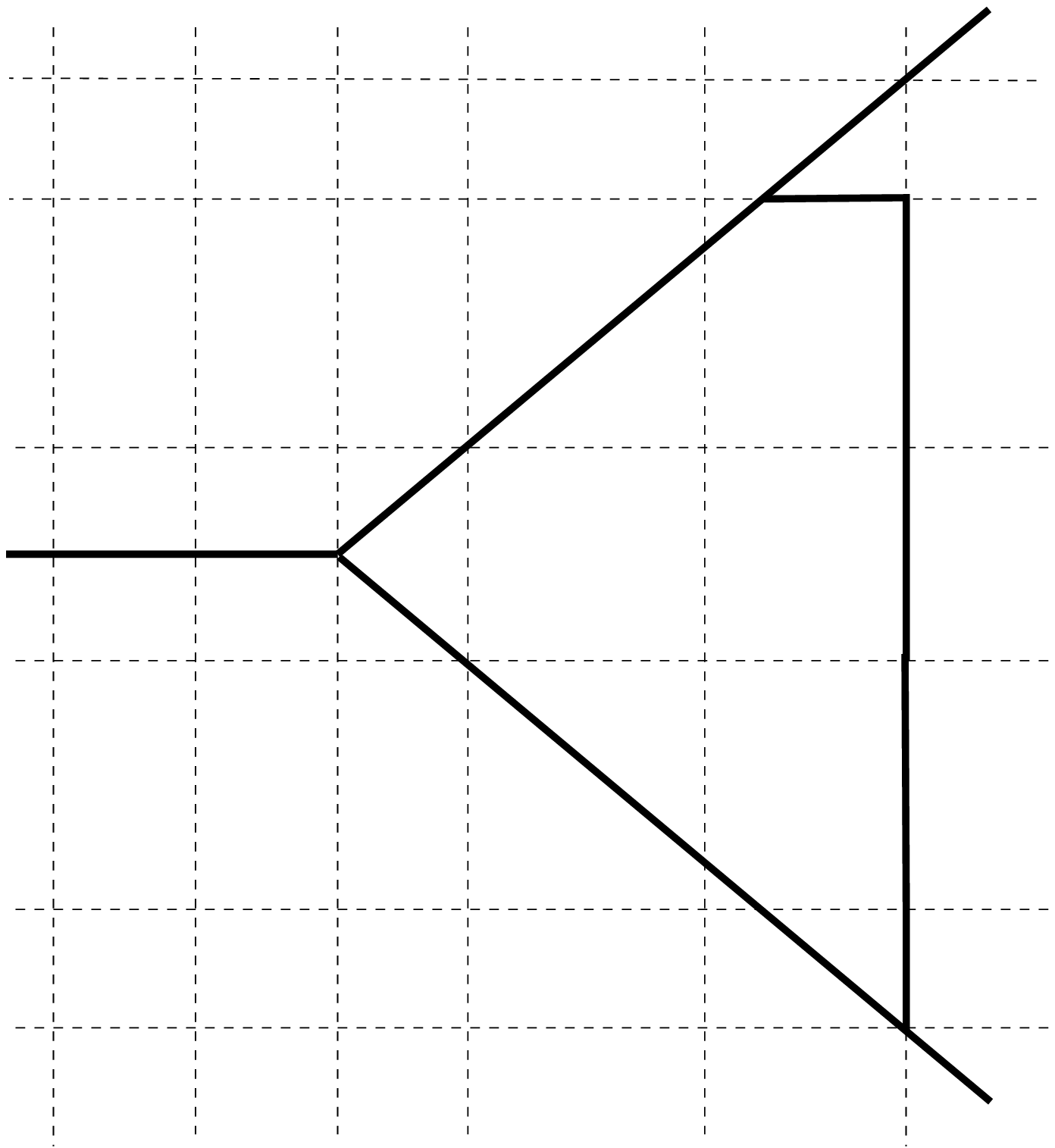}
\end{array}
\begin{array}{c}
\includegraphics[height=2.5cm]{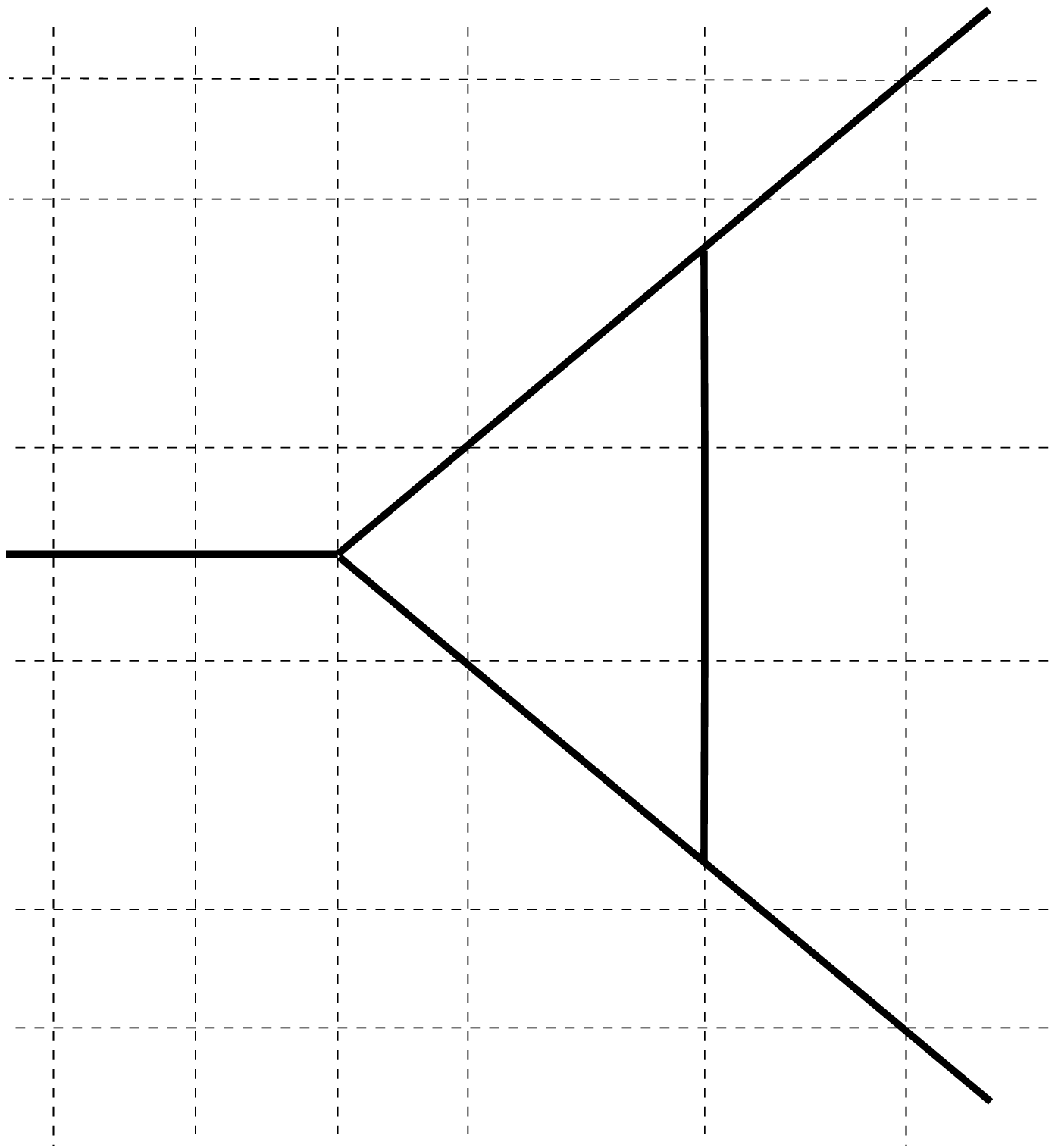}\\
\includegraphics[height=2.5cm]{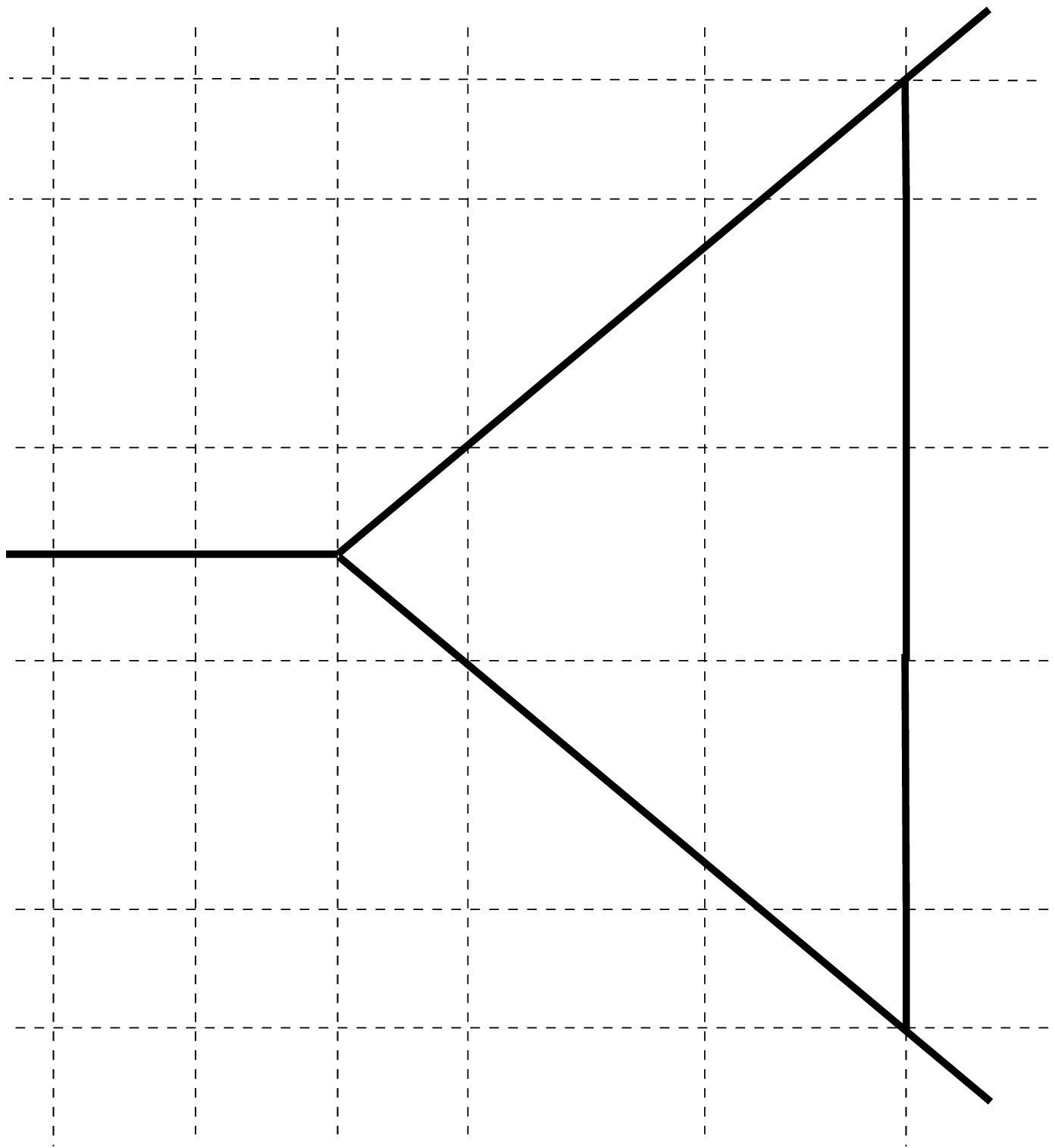}
\end{array}
\)}
\caption{A set of discrete transitions representing one of the contributing histories implied
by our regularization of the generalized projection $P$
in Equation (\ref{PS}); from left to right in two rows.}
\label{vani}
\end{figure}
\begin{figure}[h!!!!!]
\centerline{\hspace{0.5cm} \(\begin{array}{c}
\includegraphics[height=5cm]{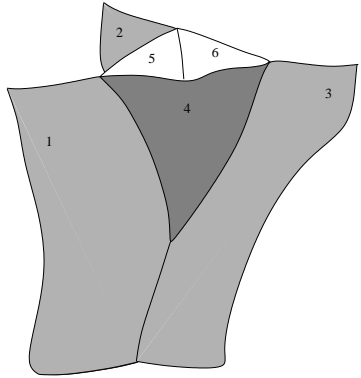}
\end{array}\) }
\caption{\small The regularization of the generalized projector $P$ produces
a continuous sequence of transitions through spin-network states that can be pictured in the form of a
continuous 2-complex. }
\label{pilin}
\end{figure}


\subsection*{3.2. Physical Hilbert space}

In the previous sections we have shown how to obtain the spin foam
representation of the physical inner product. We have also shown that our
regularization of the generalized projector $P$ implies the
Ponzano-Regge amplitudes automatically in its gauge fixed form in the language
of reference \cite{frei9}. In this section we sum up the contributions of
the infinite-many spin foams and provide a close formula for the physical
inner product for an arbitrary manifold of topology ${\cal M}=\Sigma \times \R$  where $\Sigma$ is
any Riemann surface. We also provide a basis of the physical Hilbert
space.

\subsubsection*{3.2.1. Physical scalar product}
\begin{figure}[h!!!!!]
 \centerline{\hspace{0.5cm} \( \sum \limits_{j k} \Delta_j \Delta_k
\begin{array}{c}
\includegraphics[height=3cm]{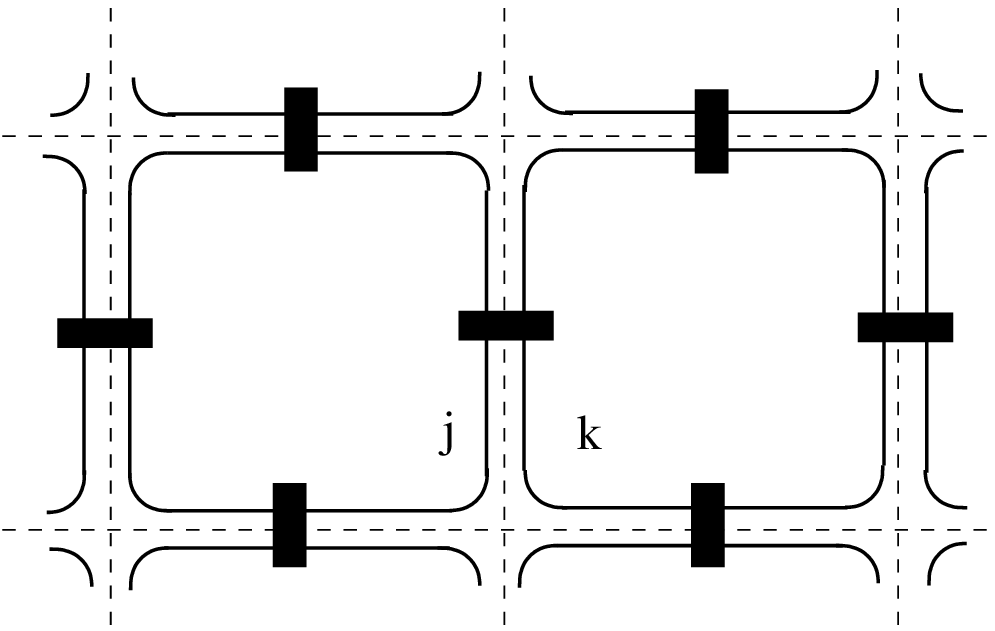}
\end{array} =  \sum \limits_k \Delta_k
\begin{array}{c}
\includegraphics[height=3cm]{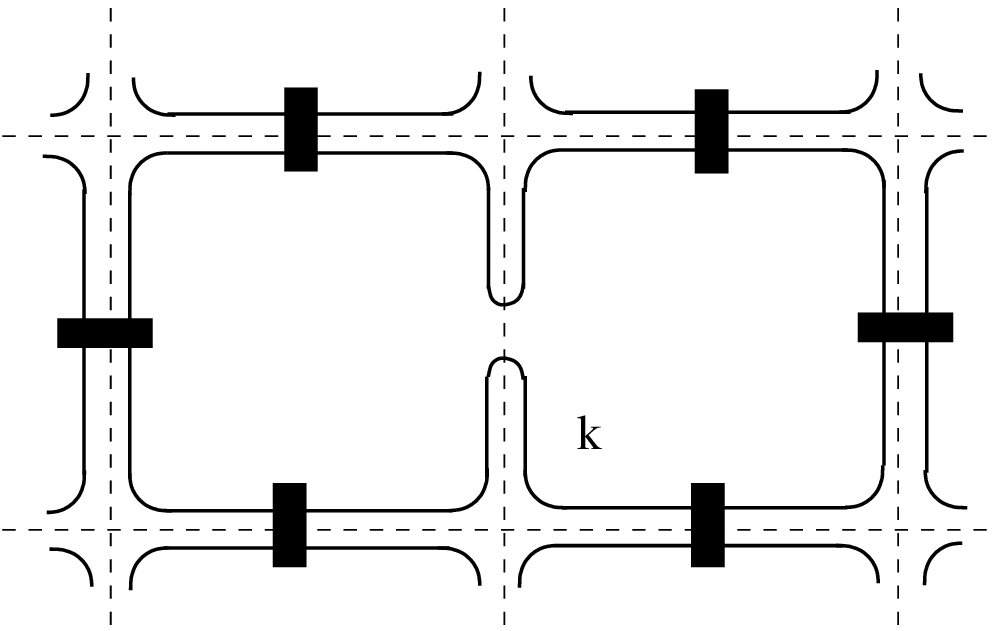}
\end{array}
\) }
\caption{\small Infinitesimal plaquette-delta-distributions can be
  integrated over to the action of a single irreducible loop (see
  equation (\ref{kaka}) in the appendix). }
\label{hh}
\end{figure}

In the computation of $\ssp$ the product of delta functions evaluated on the
holonomy corresponding to infinitesimal plaquettes that do not intersect the
graphs corresponding to $s$ and $s^{\prime}$ can be integrated over to a
single delta distribution that involves the holonomy around an irreducible
loop. This is illustrated in Figure \ref{hh} where by the integration over the
generalized connection on the common boundary of two adjacent plaquettes we
can fusion the action of the corresponding two delta distributions
into a single delta function associated to the union of the two
plaquettes. We can continue this process in the computation of $\ssp$ until we
reach the plaquettes that intersect the spin network states $s$ and $s^{\prime}$

We state the result in a more precise way. Given two spin network states
$s,s^{\prime}\in \Hk$ defined on the graphs $\Gamma_{s}$ and $\Gamma_{s^{\prime}}$
respectively; let $\Gamma_{ss^{\prime}}$ be the graph that satisfies the
following properties:
\begin{enumerate}
\item $\Gamma_{s},\Gamma_{s^{\prime}} \subset
\Gamma_{ss^{\prime}}$,
\item The graph $\Gamma_{ss^{\prime}}$ is the $1$-skeleton of a cellular decomposition
$K_{ss^{\prime}}$ of $\Sigma$,
\item The $2$-complex $K_{ss^{\prime}}$ is {\em minimal}
in the sense that it has the minimal number of $2$-cells.
We define the set of {\em irreducible loops} $\alpha^{s s^{\prime}}$ as
the set of oriented boundaries of the
corresponding $2$-cells in $K_{ss^{\prime}}$.
\end{enumerate}
With these definitions, the argument above shows that
\ba \label{irrl}  \nonumber
\ssp &:=& \ \lim_{\epsilon\rightarrow 0} \ \frac{1}{N^{\epsilon}_p!} \sum_{\sigma(\{i\})} \
\ <\prod_{p^{\sigma(i)}} \ {\delta}({U}_{p^{\sigma(i)}}) s, \ s^{\prime}>\\
\ &=&  <\prod \limits_{\gamma \in {\alpha^{s s^{\prime}}}} \
{\delta}({U}_{\gamma}) s, \ s^{\prime}>,
\ea
where ${U}_{\gamma}$ denotes as usual the holonomy around an
irreducible loop $\gamma \in {\alpha^{s s^{\prime}}}$.

\subsubsection*{3.2.2. Diffeomorphism invariance}

Using this we can explicitly see how the generalized projector $P$
implements diffeomorphism invariance or more precisely homeomorphism
invariance with our regularization. We can explicitly illustrate this with
two simple examples which sketch the general idea which can easily extended
to a proof.

\begin{figure}[h!!!!]
\centerline{\hspace{0.5cm} \(
\begin{array}{c}
\includegraphics[width=5cm]{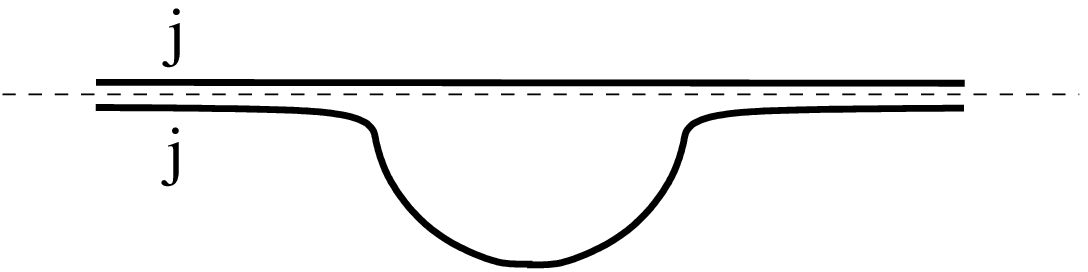}
\end{array}
\begin{array}{c}
\includegraphics[width=4cm]{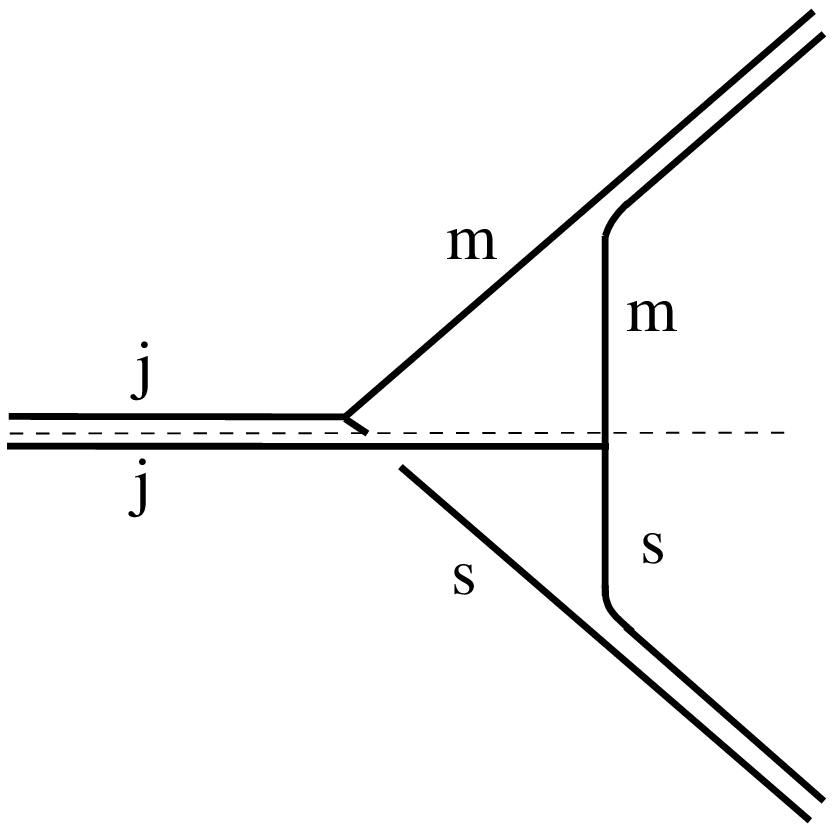}
\end{array}\) }
\caption{{\small Examples of diffeomorphism equivalent (piece of) spin-networks. On the left, the diffeomorphism acts on the edge; on the right, it acts on the vertex.}}
\label{diff}
\end{figure}

First consider a spin network and its deformation by a diffeomorphism which { acts}
trivially everywhere except for a region intersecting a single edge. The two
spin networks are orthogonal in $\Hk$. Their difference is represented on the
left of Figure \ref{diff}. Now, it follows from our definition of the physical
inner product that the transition amplitude between the two spin network
states is equal to one, namely
\ba
\nonumber &&
 \centerline{\hspace{.5cm} \(
\sum \limits_k \Delta_k \begin{array}{c}
\includegraphics[width=4cm]{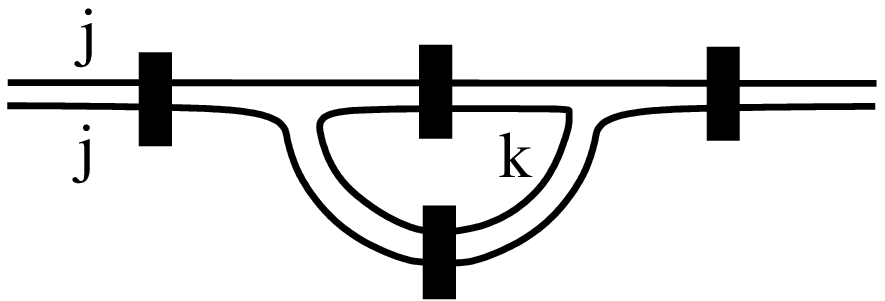}
\end{array}
=\sum \limits_k \Delta_k
\begin{array}{c}
\includegraphics[width=4cm]{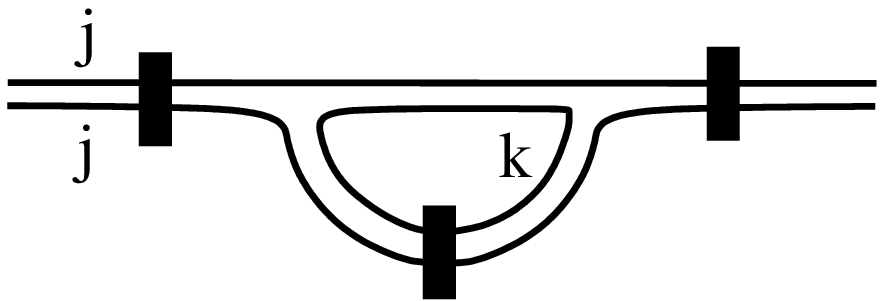}
\end{array}\) } \\ &&  \centerline{\hspace{.5cm} \(
= \begin{array}{c}
\includegraphics[width=4cm]{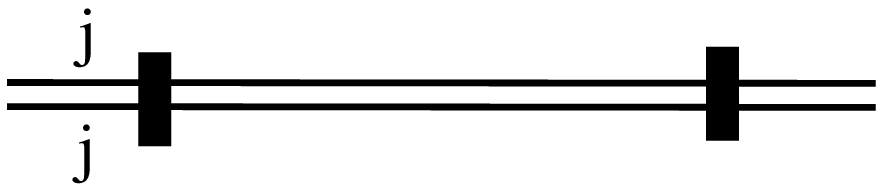}
\end{array},
\) }
\label{bb}
\ea
where in the first equality we have used the {\em gauge fixing}
identity (\ref{gfi}), and in the second we have used the {\em
summation } identity (\ref{si}). This shows that the deformed and
undeformed states are physically equivalent and the equivalence is
implemented by the operator $P$. More precisely, in the spin foam
representation we will have a continuous transition between the
initial spin network and the final one (obtained by the action of a diffeomorphism on the original one)
with transition amplitude equals one.

In order to prove that any two spin networks belonging to the same homotopy
class are physically equivalent we have to consider homeomorphisms that have
non trivial action on spin network nodes. An example of this in the case of a
3-valent node is presented on the right of Figure \ref{diff}. Again, the
direct computation of the transition amplitude shows that the initial and
final states are physically equivalent. Namely:
\ba \nonumber &&
 \centerline{\hspace{.5cm} \(
 \sum \limits_{k,p} \Delta_p \Delta_k
\begin{array}{c}
\includegraphics[width=4cm]{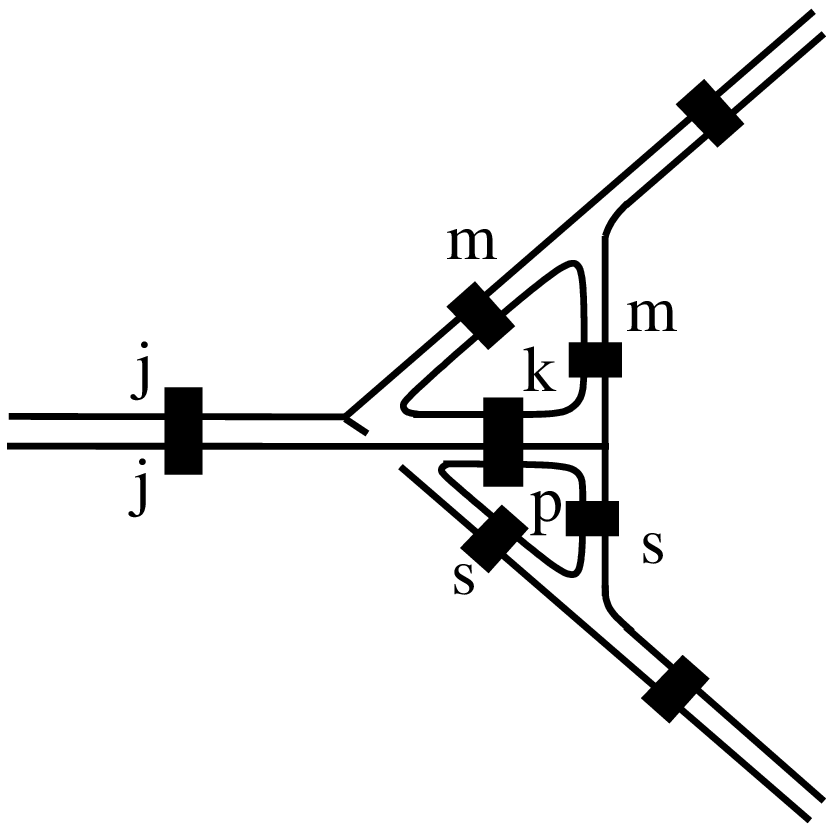}
\end{array}
 =\sum \limits_{k,p} \Delta_p \Delta_k
\begin{array}{c}
\includegraphics[width=4cm]{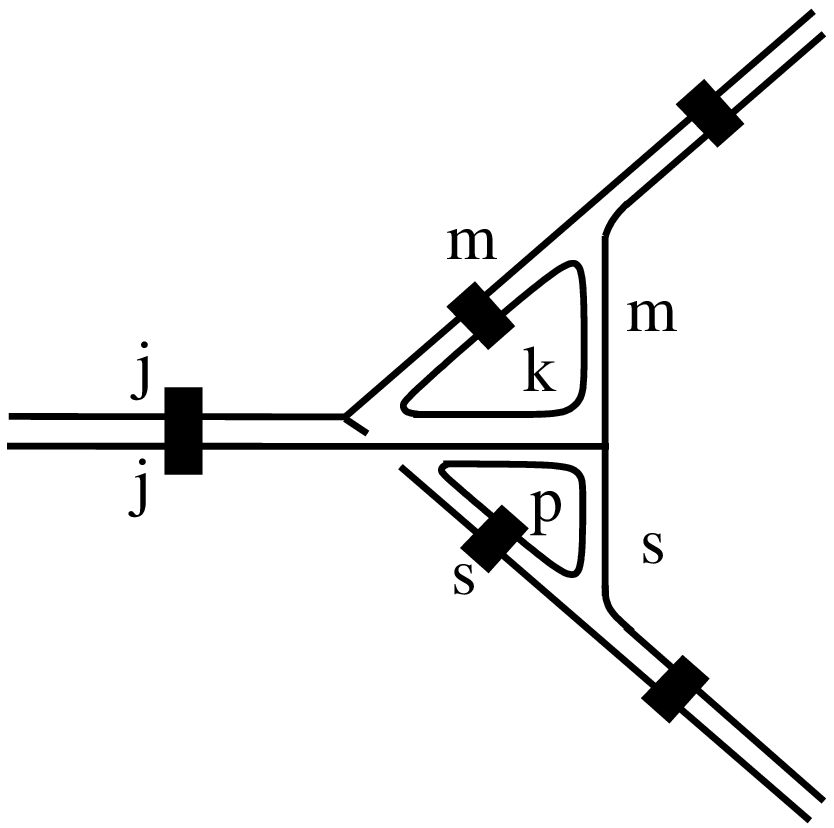}
\end{array}\)}\\ &&
 \centerline{\hspace{.5cm} \( =
\begin{array}{c}
\includegraphics[width=4cm]{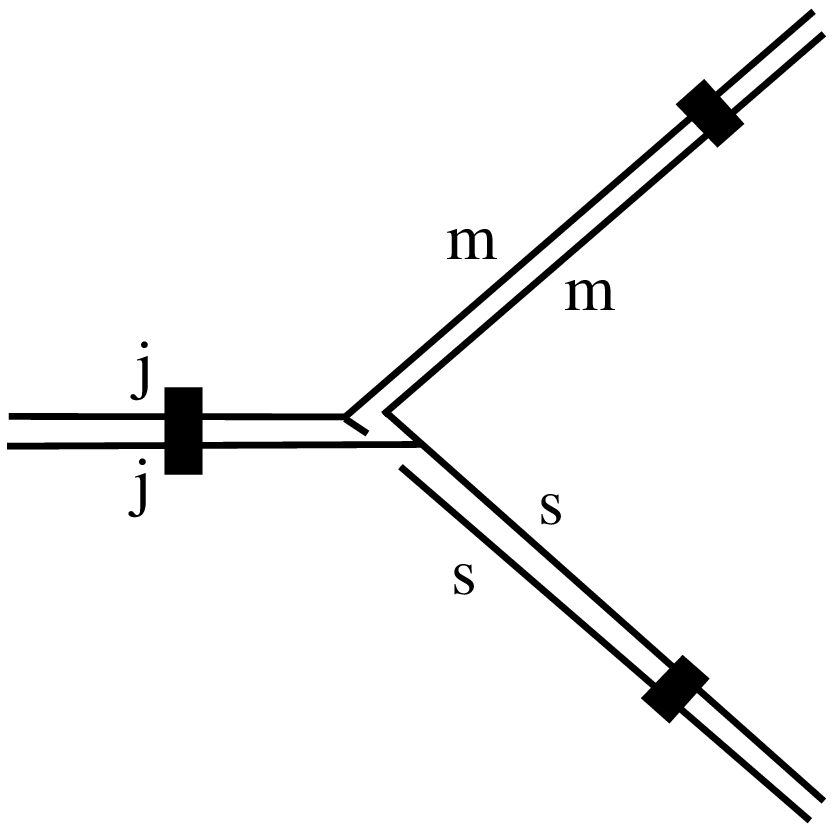}
\end{array},
\) }
\label{aa}
\ea
where again we have used the gauge fixing and summation identities.
Combining the results of (\ref{bb}) and (\ref{aa}) we can prove that any two
spin network states belonging to the same homotopy class are physically
equivalent. At the classical level we had already seen that the curvature
constraint (\ref{gauss}) generates both diffeomorphism transformations in
$\Sigma$ in addition to time reparametrization that is represented by a
fiducial notion of dynamics in the spin foam representation. We have seen how
diffeomorphism invariance is imposed by the generalized projection $P$ making
elements of the homotopy class of spin networks physically equivalent.


\subsubsection*{3.2.3. Relation with the Ponzano-Regge model}

The rest of the equivalence is represented by the well known skein relations
that relate physically equivalent spin network states in three dimensions which can be easily proved using our formalism. In three
dimensions these relations imply that any spin network state is physically
equivalent to some linear combination of spin network states based on certain
minimal graphs which do not contain any irreducible loops. One of such
skein relation follows directly from the transition amplitude corresponding to
the process shown in Figure \ref{pilin}.

In the spin foam representation we have shown that such process can be
represented by a sum over continuous spin foams as the one depicted in Figure
\ref{pilin}. However, we have not yet computed its amplitude. Using the
prescription of the irreducible loops the amplitude is simply given by the following:
\ba\nonumber && \centerline{\hspace{0.5cm} \( \Delta_1 \Delta_2 \Delta_3
\sqrt{\Delta_4 \Delta_5 \Delta_6} \sum \limits_k \Delta_k \begin{array}{c}
\includegraphics[height=3cm]{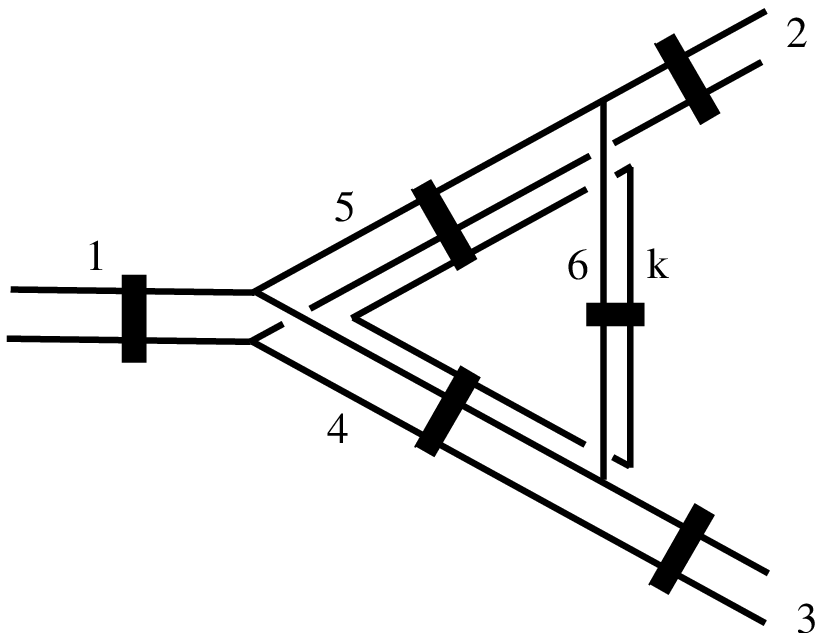}
\end{array}\)} \\ && \nonumber \centerline{\hspace{0.5cm} \( =\ \sqrt{\Delta_4 \Delta_5 \Delta_6}
\begin{array}{c}
\includegraphics[height=3cm]{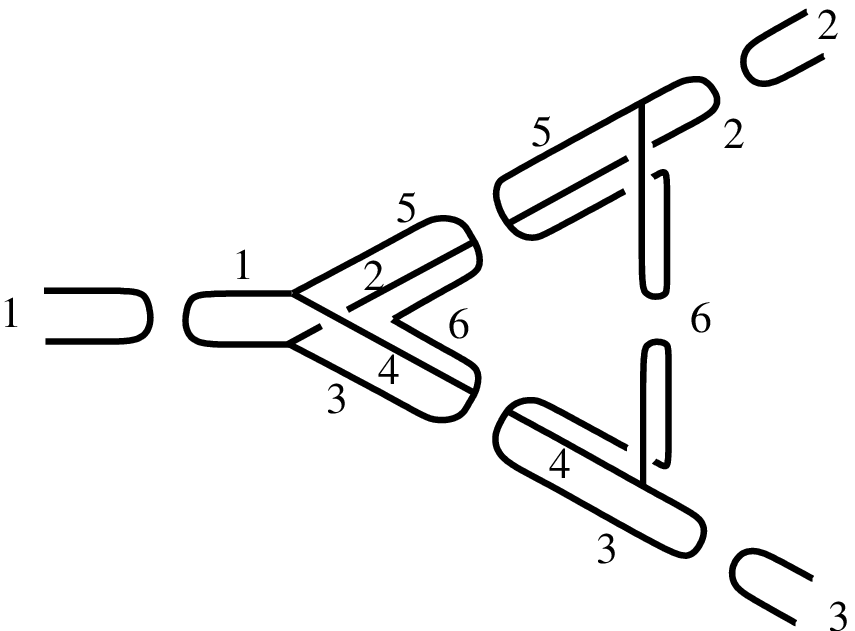}
\end{array}\)} \\ &&  \centerline{\hspace{0.5cm} \( =\sqrt{\Delta_4 \Delta_5 \Delta_6}
\left\{\begin{array}{ccc}1\ \ 2 \ \ 3\\ 4\ \  5\ \  6  \end{array}\right\}
\begin{array}{c}
\includegraphics[height=1.5cm]{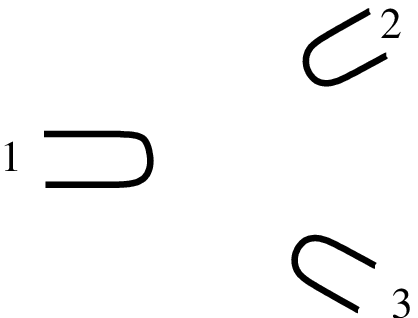}
\end{array},\) }
\label{sp}
\ea where the tetrahedron represents the contraction of four $3$-intertwiners
which defines a $6j$-symbol and the factor $\Delta_1 \Delta_2 \Delta_3
\sqrt{\Delta_4 \Delta_5 \Delta_6}$ comes from the normalization factors $
\sqrt{\Delta_4 \Delta_5 \Delta_6}$ and $\sqrt{\Delta_1 \Delta_2 \Delta_3
\Delta_4 \Delta_5 \Delta_6}$ in $\Hk$ of the corresponding spin network
states. The final amplitude on the right of the previous equation is clearly
the Ponzano-Regge amplitude for that transition: in the spin foam picture the
$6j$-symbol corresponds to the vertex amplitude and $\sqrt{\Delta_4 \Delta_5
\Delta_6}$ to the amplitude of the three new faces. In a similar way one can
prove the re-coupling identity \ba \centerline{\hspace{0.5cm} \( \Delta_1
\Delta_2 \Delta_3 \Delta_4 \sqrt{\Delta_5\Delta_6} \sum \limits_k \Delta_k
\!\!\!\!\!\begin{array}{c} \includegraphics[height=3cm]{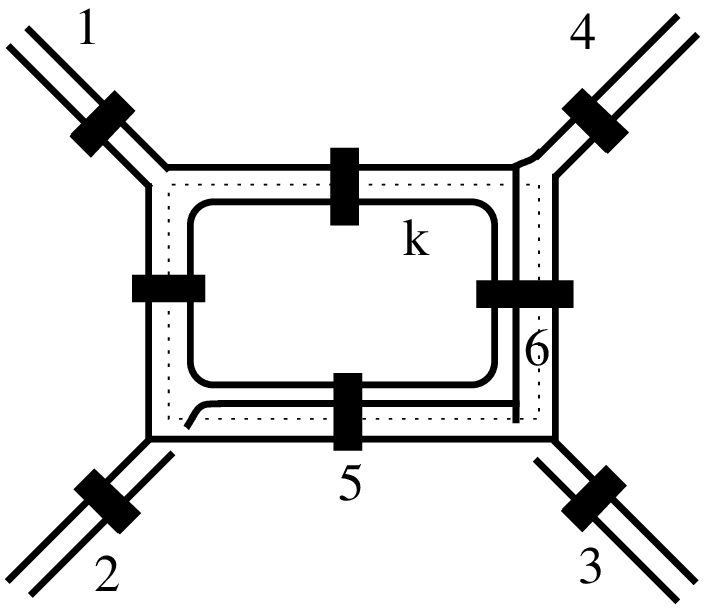}
\end{array}\!\!\!\!\!\!\!\!\!\! = \ \sqrt{\Delta_5 \Delta_6}
\begin{array}{c}
\includegraphics[height=3cm]{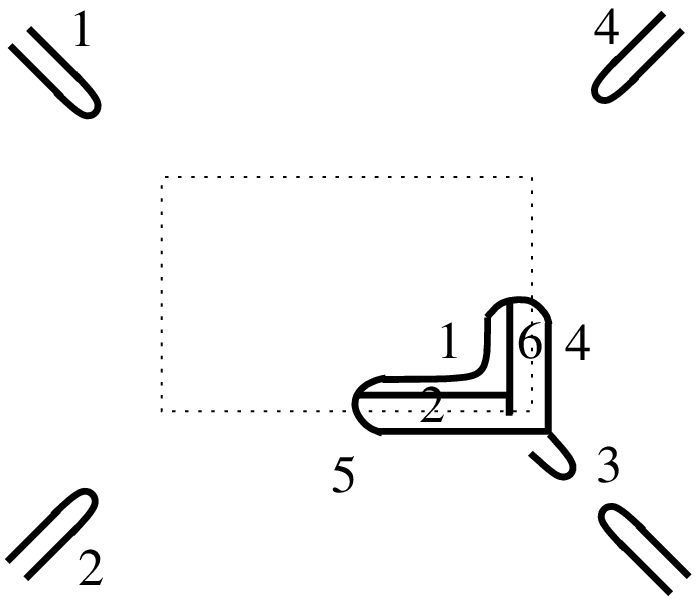},
\end{array}\)}
\label{as}
\ea
where amplitude on the right is again a $6j$-symbol.

We can write Equations (\ref{sp}) and (\ref{as}) as
\ba \nonumber &&
 \centerline{\hspace{.5cm} \({\left<\!\!\!\begin{array}{c}
\includegraphics[height=.8cm]{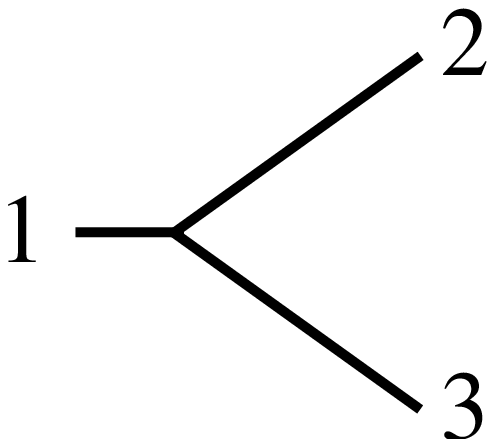}
\end{array},\!\!\!\begin{array}{c}
\includegraphics[height=.8cm]{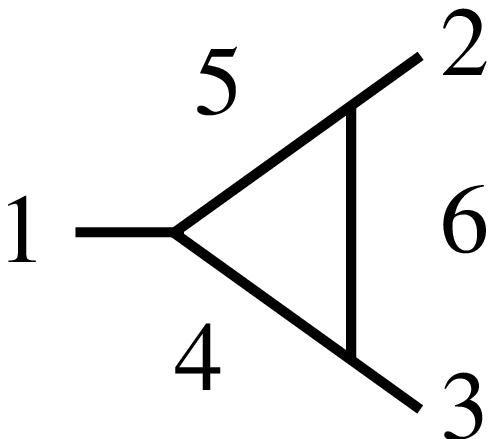}
\end{array}\right>_{\va ph}}\!\! =  \sqrt{\Delta_4\Delta_5\Delta_6}
\left\{\begin{array}{ccc}\vani 1\ \ 2 \ \ 3\\ \vani  4\ \  5\ \  6  \end{array}\right\}
\times {\rm rest}
\)} \\ &&
\centerline{\hspace{.5cm} \({\left<\!\!\!\begin{array}{c}
\includegraphics[height=.7cm]{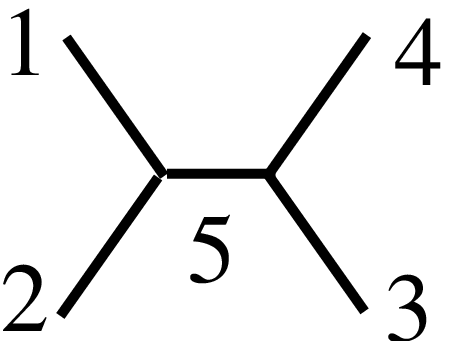}
\end{array},\!\!\!\begin{array}{c}
\includegraphics[height=.8cm]{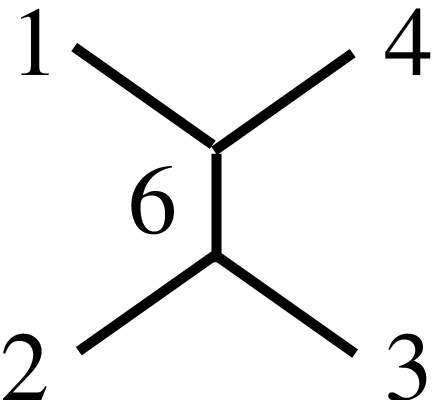}
\end{array}\right>_{\va ph}}\!\!=  \sqrt{\Delta_5\Delta_6}
\left\{\begin{array}{ccc}\vani 1\ \ 2 \ \ 3\\ \vani  4\ \  5\ \  6  \end{array}\right\}
\times {\rm rest},
\)}
\label{pachner}
\ea
Were by `{\em rest}' we denote the amplitude that follows from the details of
the rest of the states at hand.
The previous transitions correspond to the two Pachner moves that relate
any two dual of simplicial decompositions in two dimensions.
Any 2-complex dual to a triangulation of $M=\Sigma\times \R$ in three
dimensions can be foliated by a series of graphs that are dual to a
triangulation of $\Sigma$. Therefore, we can reconstruct the amplitude of any $2$-complex dual
to a three dimensional triangulation which results in the well known
Ponzano-Regge amplitude \cite{oo,arns,DePietri:1999cp}.

In our case there is an infinite series of skein relations as our states are
based on arbitrary graphs when the regulator is removed.

\subsubsection*{3.2.4. Construction physical Hilbert space}

At this stage we can give an explicit definition of the physical Hilbert space
for any space time of topology $M=\Sigma \times \R$ where $\Sigma$ is an
arbitrary Riemann surface of genus $g>1$. For the moment we will exclude the case
$\Sigma=T^2$ which will be analyzed later (the case of the sphere has been treated in Section 3.1.).
The spin foam representation of the projection operator introduced in
this section allows us to select a complete basis in $\Hp$. The generalized
projection operator $P$ can be viewed as implementing the curvature constraint
by group averaging along the orbits of the constraint on the elements of $\Hk$.
In the previous subsection we have explicitly shown how this implies the
physical equivalence of spin networks in the same homotopy class.

In addition to deforming spin network states by homeomorphisms the
action of $P$ can also create or annihilate irreducible loops as
shown in Figure 5. This process can however occur only in one of
the two directions as previously emphasized: the spin foam
representation of $P$ contains only tree-like 2-complexes with no
bubbles as a result of our definition. We need to find a minimal
set of states labeling the equivalence classes of states under the
action of the curvature constraint. In order to simplify this task
we must go in the direction in which graphs are simplified by the
elimination of irreducible loops. This can be obtained by a series
of Pachner moves of the previous section that relate physical
equivalent graphs. Let us show this in more detail.

\begin{figure}[h!!!!!!!!!!!!!!!!!!!!!!!!!!!!!!!!!]
 \centerline{\hspace{0.5cm} \(
\begin{array}{c}
\includegraphics[height=3cm]{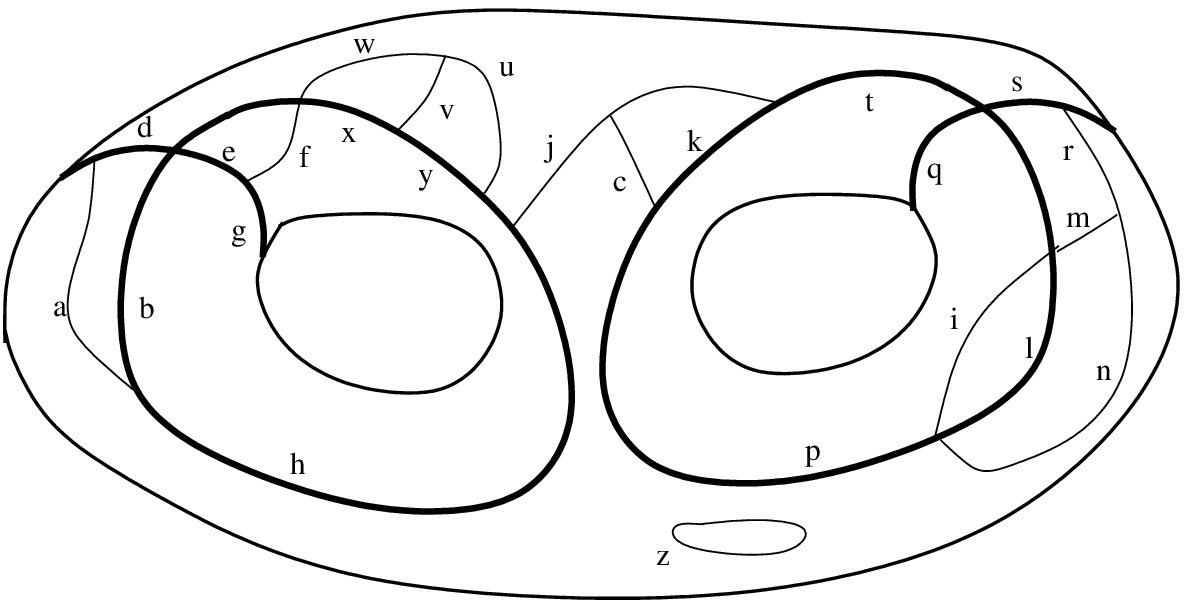}
\end{array}
\begin{array}{c}
\includegraphics[height=3cm]{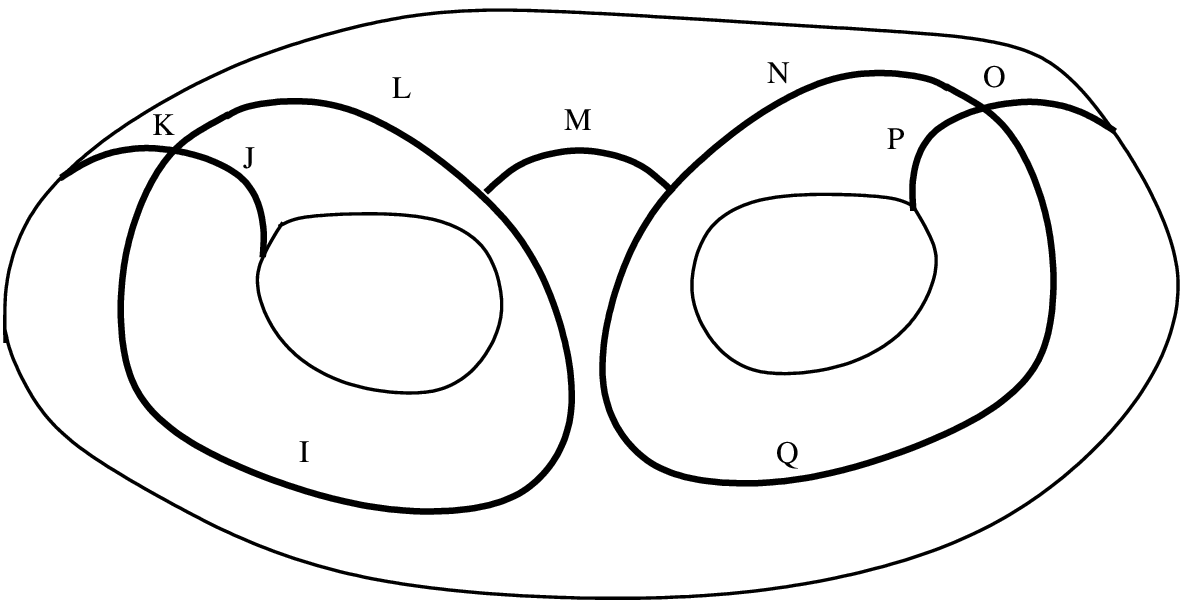}
\end{array}\) }
\caption{\small States on a genus two Riemann surface: Any arbitrary linear
combination of spin network states on the left of Figure \ref{toronja}
labeled by quantum numbers $a$ to $z$ can be `evolved' to a linear
combination of spin network states on the right (labeled by only $9$ spins).
This is however still an over-complete generators' family of $\Hp$.}
\label{toronja}
\end{figure}

Given a contractible region of an arbitrary graph we can eliminate
all the irreducible loops contained in that region by a
combination of the gauge fixing identity (\ref{gfi}) and the
summation identity (\ref{si}). As an end result all the components
of the generalized connection in that region must be set to
$\mathbbm{1}$. If we have $n$ outgoing links (labeled by
$j_1\cdots j_n$) from that region, then the state is physically
equivalent to some linear combination of the finite orthonormal
basis of intertwiners in ${\rm Inv}[j_1\otimes \cdots \otimes
j_n]$. In the spin foam picture this equivalence is represented by
an evolution from the original complicated state to the linear
combination of intertwiners by the elimination of irreducible
loops in a tree-like spin foam with no bubble. In this process we
are simply moving along the gauge orbits generated by the
curvature constraint. We can continue this process until we arrive
to a set of irreducible spin network states which in the case of
the $g=2$ Riemann surface is shown in Figure \ref{toronja}. The
generalization for higher genus is obvious. This set of
irreducible spin network states is an over-complete family of
states in $\Hp$. By construction these states label the
equivalence classes of elements $<Ps|\in Cyl^*$. In the generic
$g$ case the elements of this family are labelled by $6g-3$
quantum numbers as it can be checked.

Now we will propose a construction of a basis of $\Hp$ based on the idea
of gauge fixing the remaining gauge degrees of
freedom in the previous over complete basis.
Computing the inner product between the elements of this
over complete basis involves a single irreducible loop. It is given
by the contractible loop that can be drawn by contouring the graph
on the right in Figure (\ref{toronja}). The delta distribution
associated to that last loop imposes what is left of the curvature
constraint and is explicitly given by the following delta
function: \be \delta(g_1g_2g^{-1}_1g^{-1}_2g_3g_4g^{-1}_3g^{-1}_4
\cdots g_{2g-1}g_{2g}g^{-1}_{2g-1}g^{-1}_{2g})= \int dN {\rm
exp}(i {\rm Tr}[N U^{g}_{\ell}]),\label{putain} \ee where $g_i$
are the holonomies around the $\Pi^1(\Sigma)$ generators and
$U^{g}_{\ell}$ denotes the argument of the $\delta$-distribution.
We have re-expressed the $\delta$ as the integral of the
exponentiated constraint to present the argument that follows.
Formally the operator ${\rm exp}(i {\rm Tr}[N
  U^{g}_{\ell}])$ produces gauge transformations whose orbits are
$3$-dimensional. This tell us that in our over-complete basis there are three too
many quantum numbers. More precisely, the nine spins labeled $I,\cdots,
Q$ correspond to eigenstates of the length operators associated to paths that
are dual to the spin network graph. Clearly these operators are not Dirac
observables as they do not commute with the last constraint.

The three superfluous quantum numbers can be regarded  as
the residual gauge degrees of freedom that remain after imposing
all the local constraints up to the last global one of (\ref{putain}).
They correspond to two remaining `global' diffeomorphisms
$\Sigma$ and time reparametrization.

\begin{figure}[h!!!!!!!!!!!!!!!!!!!!!!!!!!!!!!!!!]
 \centerline{\hspace{0.5cm} \(
\begin{array}{c}
\includegraphics[height=3cm]{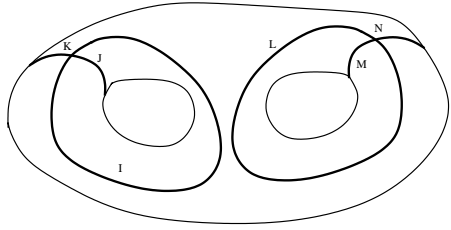}
\end{array}\) }
\caption{\small Example of a basis of $\Hp$ for the genus $g=2$ surface.}
\label{quinoto}
\end{figure}

\begin{figure}[h!!!!]
\centerline{\hspace{0.5cm} \(\begin{array}{c}
\includegraphics[height=2.7cm]{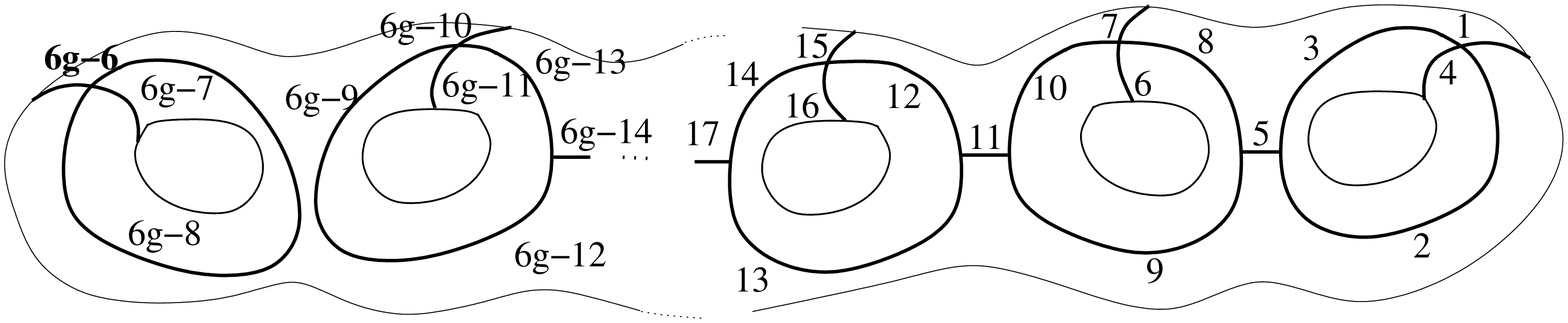}
\end{array}\) }
\caption{\small A basis of physical states for an arbitrary genus $g$ Riemann surface. Our definition of the generalized
projection $P$ directly leads to such basis of physical states.}
\label{toron}
\end{figure}

The physical Hilbert space can be characterized by the subset of spin network states obtained from the
over-complete irreducible set considered above by setting one of the spin to zero\footnote{Let us consider a simplified version of the
  situation at hand defined by a non relativistic particle confined to a sphere
  with the constraint $\hat \phi=0$. The auxiliary Hilbert
  space is taken to be the space of continuous functions on the
  sphere. Namely, a general state can be represented by a wave function
\[\Psi(\theta,\phi)=\sum_{\ell, m} c^{\ell m} Y_{\ell m}(\theta, \phi)\]
where $Y_{\ell m}(\theta,\phi) = P_\ell^m(cos(\theta)) e^{im\phi}$ are spherical harmonics expressed in terms of the Lagrange polynomials $P_\ell^m$.
The physical scalar product
\[<\Psi,\Phi>_{phys}=<\Psi,\sum_n e^{i n \hat \phi}\Phi>=
\int d\Omega \delta(\phi) \overline{ \Psi(\theta,\phi)} \Phi(\theta,\phi)=
\int d[{\rm cos}(\theta)] \overline{\Psi(\theta,0)} \Phi(\theta,0).\]
In terms of the elements of the $Y_{\ell m}$ basis of $\Hk$---the analog
of the spin network states in gravity---the physical inner product takes the form
\[<\ \ell\ m,\ \ell^{\prime}\ m^{\prime}>_{phys}= \sqrt{\frac{(2\ell+1)(2\ell^{\prime}+1)}{16 \pi^2}\frac{(\ell-m)!(\ell^{\prime}-m^{\prime})!}
{(\ell+m)!(\ell^{\prime}+m^{\prime})!}}\ \ \int \limits_{-1}^{1}dx
\ P_{\ell}^{m}(x)P_{\ell^{\prime}}^{m^{\prime}}(x),\] where we
have made the substitution $x={\rm cos}(\theta)$ and we have
introduced the standard ket notation $|\ell,m>$ to denote the
quantum states. As in the gravity case the kinematical orthogonal
states are no longer orthogonal as it can be easily checked using
the previous equation. The states $|\ \ell\ m>$ are an over
complete basis of $\Hp$. A complete basis can be obtained if we
fix the value of the $\hat L_z$ operator (conjugate to the
constraint) to the value $0$, i.e. $m=0$. Moreover, for any
$(m,\ell)$, we can always expand $P_{\ell}^m$ in terms of
$P_\ell^0$ as follows:
\[P_{\ell}^{m}(x)=\sum_{\ell^{\prime}} \alpha^{\ell^{\prime}}_{\ell,m} \ P_{\ell^{\prime}}^{0}(x).\]
Finally, this implies that
\[|\ \ell\  m> \phys \sum_{\ell^{\prime}} \alpha^{\ell^{\prime}}_{\ell,m}\  \ |\ \ell^{\prime}\  0>\]
so that the $\{|\ \ell \ 0>\}$ can be taken as a basis of $\Hp$.
}
(e.g.,  $M=0$ in Figure (\ref{toronja})). Notice that the new set of states
are labeled by precisely $6g-6$ quantum numbers corresponding to the
expected number of physical degrees of freedom (in Figure (\ref{quinoto}) we
illustrate the case $g=2$; the case $g>2$ is illustrated in Figure
(\ref{toron})). One could consider the fixing of the spin of the intermediate link to a value
different from zero. That would correspond however to a partial fixation as only the magnitude
of $E$ is fixed. This can be seen in the fact that unphysical configuration remain represented by the finite
set of possible spin labels for the adjacent links corresponding to the quantized directions for the operator
$E$.


Notice that the number of
quantum numbers is $6g-6$ corresponding to the dimension of the moduli space
of $SU(2)$ flat connections on a Riemann surface of genus $g$. In this way we
arrive at a fully combinatorial definition of the standard $\Hp$
by reducing the infinite degrees of freedom of the kinematical phase space to
finitely many by the action of the generalized projection operator $P$.
The result coincides with the one obtained by
other methods that explicitly use the fact that the reduced phase space is
finite-dimensional.

We must point out that one has still to show that the elements of
the previous family are independent. We however think from the gauge fixing argument
given above that this is in fact the case.

\begin{figure}[h!!!!!!!!!!!!!!!!!!!!!!!!!!!!!!!!!!!!!!!!!!]
\centerline{\hspace{0.5cm} \(\begin{array}{c}
\includegraphics[height=3cm]{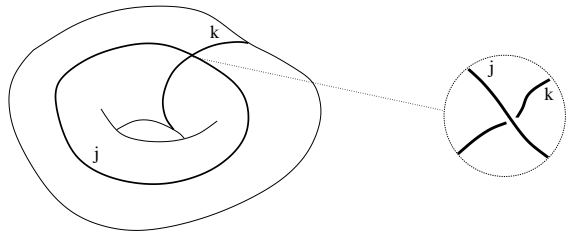}
\end{array}\) }
\caption{\small A basis of physical states in the case of the torus ($g=1$).}
\label{toroni}
\end{figure}

The case of the Torus ($g=1$) is singular. It is however easy to show that
the basis of the physical Hilbert space is generated, as a vector space, by
the two loops as shown in Figure (\ref{toroni}). The physical inner
product defined in (\ref{final1}) is divergent and needs regularization.
This problem is well known and has been investigated in the literature using reduced phase
space quantization\cite{Ashtekar:1989qd}.

\section*{4. Conclusion and discussion}

Our paper establishes a clearcut connection between the canonical
formulation of loop quantum gravity in three dimensions and previous
covariant path integral definitions of the quantum theory. The
Ponzano-Regge model amplitudes are recovered from the Hamiltonian
theory and its `{\em continuum limit}' in the sense of Zapata
\cite{za1} is built in from the starting point. Divergences that plague
the standard definition of spin foam models are not present and the
formalism provides a clear understanding of their origin which is
complementary to the covariant analysis provided in \cite{frei9}. It also
provides an explicit realization of Rovelli's proposal
for resolving dynamics in loop quantum gravity.

We would like to emphasize the fact that we are getting fully
background independent description of the spin foam in this case. The
generalized projection operator $P$, providing spin foam amplitudes,
is well defined independently of any background structure (such as a
space time triangulation) and directly from the Hamiltonian
picture. Our result is fully consistent with the standard formulation
based on the quantization of the reduced phase space.

The simplicity of the reduce phase space quantization in 2+1 gravity
might make reluctant to adopt the view point explored here. However,
the reduced phase space quantization does not seem viable in 3+1
gravity. The spin foam perspective, fully realized here in 2+1
gravity, can bring new breath to the problem of dynamics in 3+1
gravity. Needless to say that the challenges in going to 3+1 are many
and certainly no straightforward generalization of this work should be
expected. In reference \cite{Mikovic:2004im} an application of similar techniques is
used to investigate a `flat' solution to the constraints of Riemannian
3+1 gravity.


Generalization to BF theory in any dimensions seems straightforward.
The new feature in higher dimensions is the fact that the Bianchi identity
plays an important role in the regularization. If the dimension of $\Sigma$ is
larger or equal to $3$ then the curvature constraints $F=0$ are no longer
independent since we also have that $dF=0$. Through the non-Abelian Stokes
theorem this implies that certain order integration of $F$ on $2$-dimensional
closed surfaces vanishes. The regularization of $P$, introduced in Equation
(\ref{PS}), must then be modified in order to avoid the inclusion of redundant
delta functions.

The generalization to non vanishing cosmological constant is the natural next
step. In this case the curvature constraint becomes $F^i-\Lambda \epsilon^i{}_{jk} \; e^j\wedge e^k
=0$. If $\Lambda>0$ the theory can be quantized using the
Chern-Simons formulation \cite{Alek,buff} and the result involves the utilization of
quantum groups. The `path integral' version of the quantum theory is thought
to  correspond to the Turaev-Viro invariant of $3$-manifolds \cite{tur,TV}
which can be viewed as a generalization of the Ponzano-Regge model
based on $U_q(su(2))$ for $q={\rm exp}(i \frac{2 \pi}{k+2})$ where $k=\Lambda^{-1/2}$. It would be important to understand whether
these results can be obtained from the loop quantum gravity perspective
presented here. The question is whether there is a well defined regularization
of the projection operator $P$ in this case and whether its definition implies
the mathematical structures obtained by other means. Preliminary results indicate
that this is the case. Namely, that one can start with the kinematical Hilbert space
$\Hk$ of (classical) $SU(2)$ spin networks and that value of the spin foam amplitudes
derived from the arising from regularization of $P$ in this case can be
expressed in terms of the Turaev-Viro invariants and it generalizations \cite{a1}.

In a companion paper \cite{kayo} we use the formalism presented here to
provide the complete quantization of general relativity coupled to point
particles in three dimensions.

\subsubsection*{Aknowledgments}
We want to thank Abhay Ashtekar, Laurent Freidel, Carlo Rovelli, Rodolfo Gambini, Michael Reisenberger and Jorge Pullin for stimulating discussions. This work has been supported by NSF grants PHY-0090091 and INT-0307569 and the Eberly Research Funds of Penn State University.

\subsection*{APPENDIX: $SU(2)$ Haar measure and useful identities}

In this section we present the various identities that are used throughout the
paper. The main identities where use in the case of $SU(2)$ in \cite{a1}. For
a general treatment see \cite{oe}.
We start by recalling the graphical notation that is commonly used in dealing with
group representation theory. A spin $j$ unitary irreducible $SU(2)$ representation
matrix $\Pi^{j}(g)_{\va B}^{\va A}$ is represented by an oriented line going from the
free index $B$ and ending at $A$. The group element $g\in SU(2)$ on which the
representation matrix is evaluated is depicted as a dark dot in the middle. Namely,
\be
 \centerline{\hspace{0.5cm} \(
\Pi^{j}(g)_{\va B}^{\va A}=
\begin{array}{c}
\includegraphics[height=2cm]{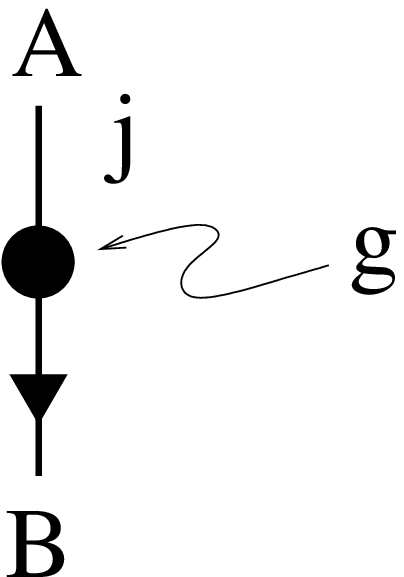}
\end{array}.\) }
\label{}
\ee
Matrix multiplication is represented by the appropriate joining
of the line endpoints, more precisely
\be
 \centerline{\hspace{0.5cm} \(
\Pi^{j}(g)_{\va B}^{\va A}\Pi^{j}(h)_{\va C}^{\va B}=
\begin{array}{c}
\includegraphics[height=2cm]{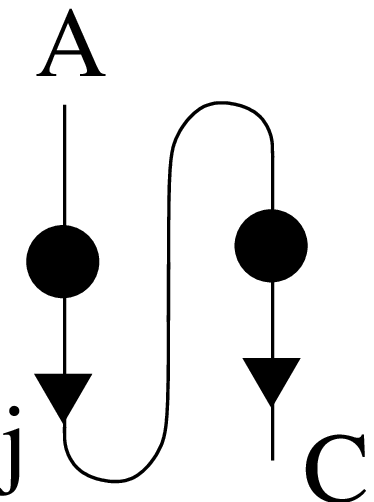}
\end{array}
=
\begin{array}{c}
\includegraphics[height=2.2cm]{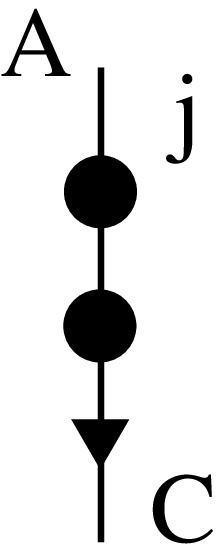}
\end{array}
.\) }
\label{}
\ee
For simplicity, we will drop the explicit reference to the
free indices at the tips of the representation lines.
The tensor product of representation matrices is
simply represented by a set of parallel lines carrying
the corresponding representation labels and orientation.
An important object is the integral of the tensor product of
unitary irreducible representations. We denote the Haar measure integration
by a dark box overlapping the different representation lines as follows:
\be
 \centerline{\hspace{0.5cm} \(
I_{B_1\cdots B_n}^{A_1\cdots A_n}=\int dg \Pi^{1}(g)_{\va B_1}^{\va A_1}\Pi^{2}(g)_{\va B_2}^{\va A_2}\cdots \Pi^{n}(g)_{\va B_n}^{\va A_n}=
\begin{array}{c}
\includegraphics[height=2cm]{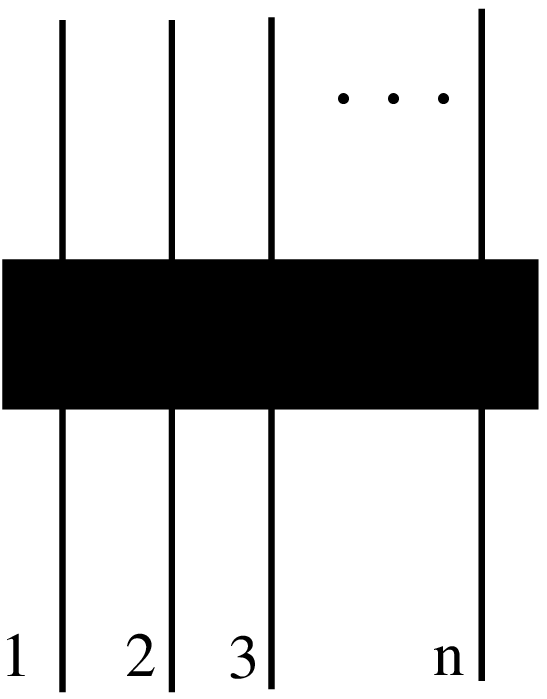}
\end{array}.\) }
\label{lulingui} \ee The invariance of the Haar measure implies
$I^2=I$ and the invariance of $I$ under right and left action of
the group; therefore, $I$ defines the projection operator $I^{1 2
\cdots n}:1 \otimes 2 \otimes \cdots \otimes n \rightarrow {\rm
Inv}[1 \otimes 2 \otimes \cdots \otimes n]$. With this in mind it
is easy to write the basic identities that follow from the
properties of the Haar measure, namely \be
 \centerline{\hspace{0.5cm} \(
\begin{array}{c}
\includegraphics[height=2cm]{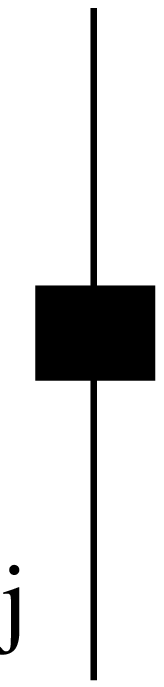}
\end{array}={ \delta_{j,0}},\) }
\label{}
\ee
\be
 \centerline{\hspace{0.5cm} \(
\begin{array}{c}
\includegraphics[height=2cm]{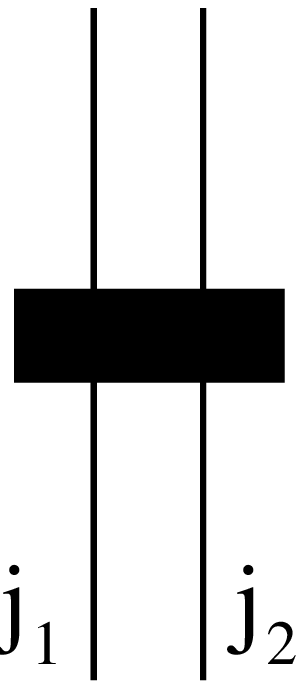}
\end{array}= \frac{1}{2j_1+1} \ \delta_{j_1 j_2}
\begin{array}{c}
\includegraphics[height=2cm]{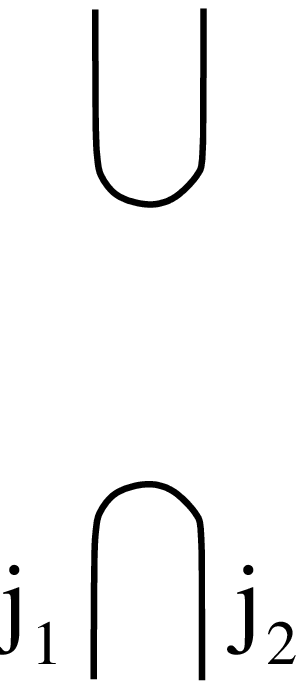}
\end{array},\) }
\label{kaka}
\ee
and generally
\be
 \centerline{\hspace{0.5cm} \(
\begin{array}{c}
\includegraphics[height=2cm]{BOX.eps}
\end{array}=\sum\limits_{\iota_1\cdots \iota_{n-3}}
\begin{array}{c}
\includegraphics[height=2cm]{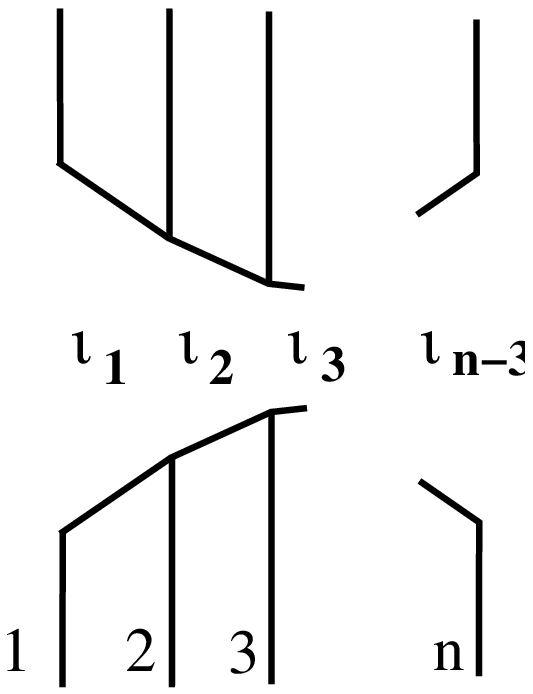}
\end{array}\) }
\label{}
\ee
where the right hand side is an expansion of the
projector operator $I^{1 2 \cdots n}:1 \otimes 2 \otimes \cdots \otimes n \rightarrow {\rm Inv}[1 \otimes 2 \otimes \cdots \otimes n]$
in terms of a normalized basis of invariant vectors in ${\rm Inv}[1 \otimes 2 \otimes \cdots \otimes n]$. In the case of $SU(2)$
the invariant vectors in such basis can be labeled by $n-3$ half integers $\iota_1 \cdots \iota_{n-3}$.

Other identities that we extensively use in this work where proved (at a more general level) in \cite{a1}.
The first is the so-called {\em summation identity}
\be
 \centerline{\hspace{0.5cm} \(
\sum \limits_k \Delta_k \ \begin{array}{c}
\includegraphics[height=2cm]{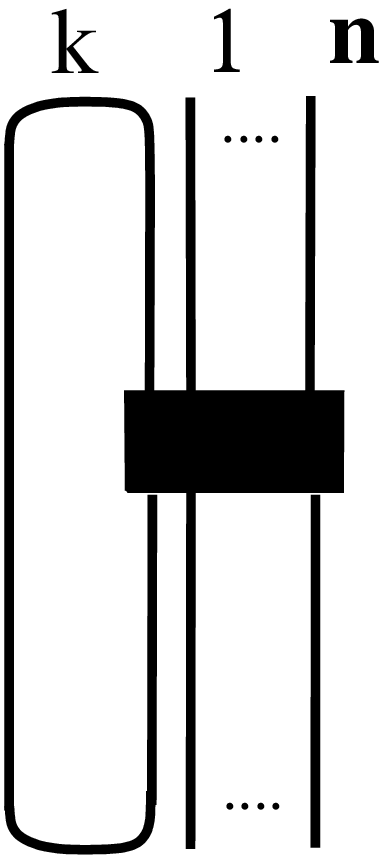}
\end{array}=
\begin{array}{c}
\includegraphics[height=2cm]{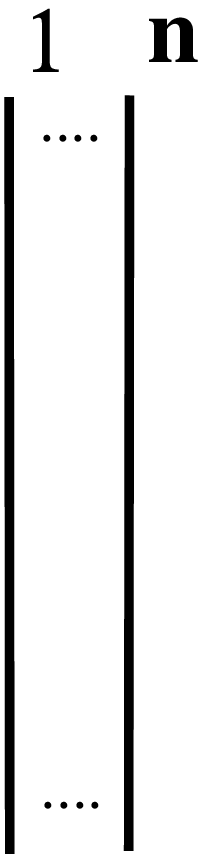}
\end{array}\) }
\label{si}
\ee
The other important identity proved in \cite{a1} is the so-called
{\em gauge fixing identity}. Given a general graph with Haar measure integrations
(represented by dark boxes) if we can draw a close loop with no self-intersections that
intersects the graph only through dark boxes, then we can erase one of the boxes without changing the evaluation.
Graphically,
\be
 \centerline{\hspace{0.5cm} \(
 \begin{array}{c}
\includegraphics[height=3cm]{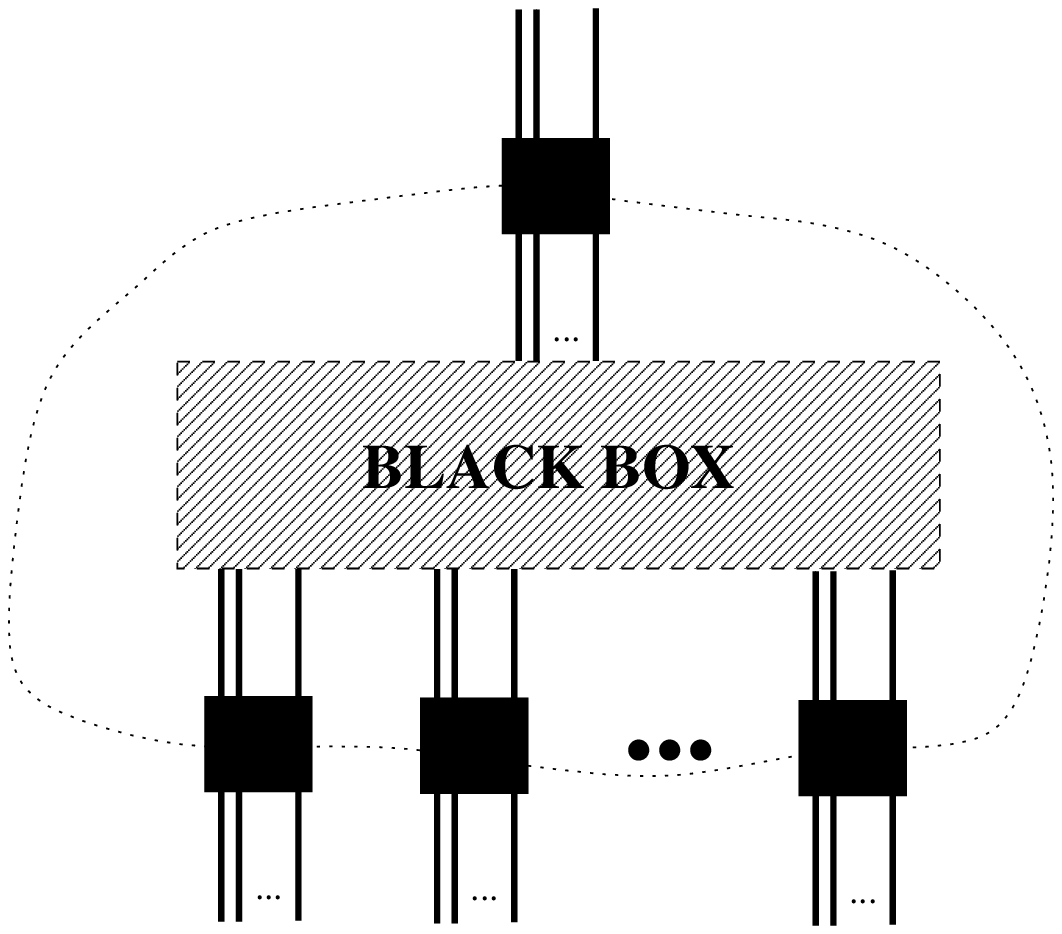}
\end{array}=
\begin{array}{c}
\includegraphics[height=3cm]{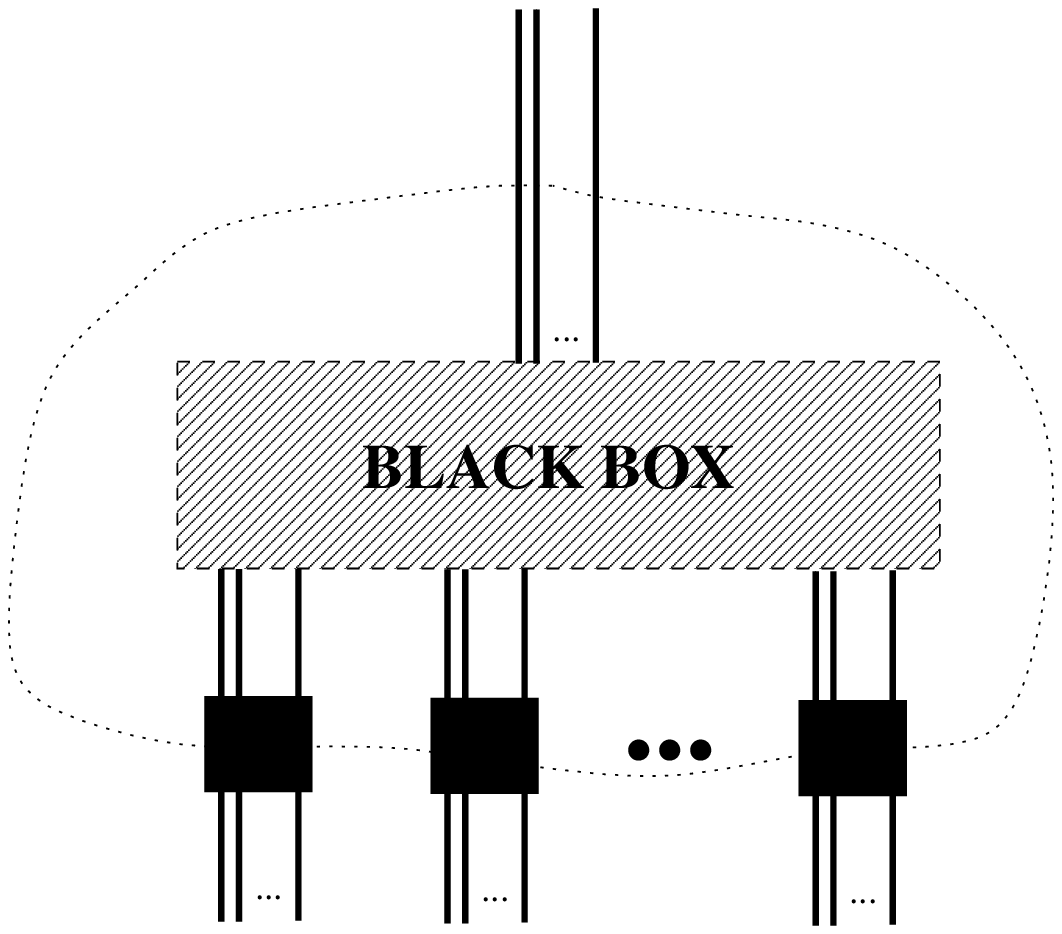}
\end{array}\) }
\label{gfi}
\ee


\end{document}